\documentclass[fleqn,usenatbib]{mnras}
\usepackage{comment}
\usepackage{newtxtext,newtxmath}

\usepackage[T1]{fontenc}

\DeclareRobustCommand{\VAN}[3]{#2}
\let\VANthebibliography\thebibliography
\def\thebibliography{\DeclareRobustCommand{\VAN}[3]{##3}\VANthebibliography}

\usepackage{graphicx}	
\usepackage{amsmath}	
\usepackage[dvipsnames]{xcolor}

\newcommand{\tratio}{\ifmmode{t_{\rm cool,mix} / t_{\rm cc}}\else{$t_{\rm cool,mix} / t_{\rm cc}$}\fi}
\newcommand{\cloudRadius}{\ifmmode{R_{\mathrm cl}}\else{$R_{\mathrm cl}$}\fi}
\newcommand{\rcl}{\ifmmode{R_{\rm cl}}\else{$R_{\rm cl}$}\fi}
\newcommand{\vrel}{\ifmmode{v_{\rm rel}}\else{$v_{\rm rel}$}\fi}

\title[Cooling \& Draping]{Better Together: The Complex Interplay Between Radiative Cooling and Magnetic Draping}

\author[Hidalgo-Pineda, Farber \& Gronke]{
Fernando Hidalgo-Pineda$,^{1,2}$\thanks{E-mail: fernando.hidalgo.pineda@gmail.com}
Ryan Jeffrey Farber$^{2}$,
Max Gronke$^{2}$
\\
$^{1}$School of Physics \& Astronomy, University of Glasgow, Glasgow G12 8QQ, UK\\
$^{2}$Max Planck Institute for Astrophysics, Karl-Schwarzschild-Str. 1, D-857481 Garching, Germany \\
}

\date{Draft from \today}

\pubyear{2023}

\begin{document}
\label{firstpage}
\pagerange{\pageref{firstpage}--\pageref{lastpage}}
\maketitle

\begin{abstract}
Rapidly outflowing cold H-I gas is ubiquitously observed to be co-spatial with a hot phase in galactic winds, yet the ablation time of cold gas by the hot phase should be much shorter than the acceleration time. Previous work showed efficient radiative cooling enables clouds to survive in hot galactic winds under certain conditions, as can magnetic fields even in purely adiabatic simulations for sufficiently small density contrasts between the wind and cloud. In this work, we study the interplay between radiative cooling and magnetic draping via three dimensional radiative magnetohydrodynamic simulations with perpendicular ambient fields and tangled internal cloud fields. We find magnetic fields decrease the critical cloud radius for survival by two orders of magnitude (i.e., to sub-pc scales) in the strongly magnetized ($\beta_{\rm wind}=1$) case. Our results show magnetic fields (i) accelerate cloud entrainment through magnetic draping, (ii) can cause faster cloud destruction in cases of inefficient radiative cooling, (iii) do not significantly suppress mass growth for efficiently cooling clouds, and, crucially, in combination with radiative cooling (iv) reduce the average overdensity by providing non-thermal pressure support of the cold gas. This substantially reduces the acceleration time compared to the destruction time (more than due to draping alone), enhancing cloud survival. Our results may help to explain the cold, tiny, rapidly outflowing cold gas observed in galactic winds and the subsequent high covering fraction of cold material in galactic halos.
\end{abstract}



\section{Introduction}
\label{sec:intro}
The formation of galaxies involves a complex interplay between cooling, heating, infall, and outflow. 
Galactic outflows in particular play a key role in the chemical and dynamical evolution of galaxies \citep[e.g.,][]{Veilleux2005,Rupke2018,zhang2018review}. 
Galactic outflows redistribute angular momentum, enabling the formation of extended disks \citep[][]{brook2011hierarchical,ubler2014stellar} and 
reorient magnetic field lines, catalyzing the growth of large-scale magnetic fields in dwarf galaxies \citep[][]{moss2017galactic}.
The mass-metallicity relation \citep[][]{lequeux1979chemical,tremonti2004origin} may be explained by galactic outflows preferentially ejecting metals from low mass halos \citep[][]{larson1974effects,Mac1999}, enriching the intergalactic medium with (observed) metal line absorbers \citep[][]{hellsten1997metal,Steidel2010,Booth2012}.
Moreover, the missing baryons problem \citep[][]{Bell2003} may be resolved by galactic outflows ejecting gas (e.g., \citealt{Mac1999}) from disks or preventing gas from accreting onto disks \citep[][]{Somerville2015,pandya2020first}, creating a vast reservoir of gas in the circumgalactic medium \citep[][]{tumlinson2005hot,werk2014cos,tumlinson2017circumgalactic,qu2022cosmic,FaucherGiguereOh_review}.

Indeed galactic outflows are detected ubiquitously: at high redshifts, in dwarf starbursts, nearby ultraluminous infrared galaxies, and in the Local Group \citep[][]{chisholm2017mass,rudie2019column,Veilleux2020,stacey2022red}. 
Classically, galactic outflows are expected to arise when clustered supernovae inject thermal energy, inflating a hot shocked over-pressurized bubble into the ambient interstellar medium \citep[][]{Chevalier1985,thompson2016origin,bustard2016versatile}. 
The injection of thermal energy by supernovae develops a wind solution analogous to stellar winds \citep[][]{weaver1977interstellar,lancaster2021efficiently}. 
X-ray observations \citep[][]{strickland2009supernova} detect hot winds in remarkable agreement with the energy loading predicted by \citet{Chevalier1985}. 
Yet galactic wind models suggest the hot phase cannot carry sufficient mass to resolve discrepancies between the observed luminosity function and the predicted halo mass function \citep[][]{Mac1999,guo2010galaxies}. 

Multiwavelength observations reveal galactic outflows are ubiquitously multiphase, containing H-I gas co-spatial with the hot ionized component and outflowing at one-few of the escape velocity of the system \citep[e.g.,][]{heckman1990nature}. 
Quasar absorption-line studies detect cold 10$^{4-5}$\,K gas in the CGM of the Galaxy, as additionally measured in emission with H-I and at colder temperatures with CO \citep[][]{putman2002hipass,DiTeodoro2018,Fox2019,su2021molecular}. 
A larger body of literature has focused in detail on the multiphase gas dynamics in the winds \citep[e.g.][]{schneider2018introducing,debora2023,li2020simple} and the impact of the launching mechanisms such as supernova feedback \citep[][]{martizzi2016supernova, gatto2017silcc,fielding2018clustered}, stellar winds
\citep[][]{kim2018numerical} or cosmic rays \citep[]{girichidis2016silcc,pakmor2016galactic,simpson2016role,ruszkowski2017global,Farber2018}. Recent work has studied in both interstellar medium (ISM) patch simulations \citep{holguin2019role,rathjen2021silcc,girichidis2022spectrally,habegger2022impact} as well as global models \citep{lita2021synthetic,pandya2021characterizing,steinwandel2022structure,steinwandel2022driving,farber2022stress} the structure and evolution of winds which either fountain flow or eject a cold phase. Despite the progress in full wind simulations \citep[]{jacob2018dependence,hopkins2021cosmic}
(for earlier work see \citealt{naab2017theoretical} and references therein), 
much work remains to disentangle the relevance of the competing processes in driving galactic winds and reproducing in detail observations of multiphase, mass-loaded outflows.

For the simplest picture of a multiphase, galactic wind, consider a cold cloud of density $\rho_{\rm c}$ in pressure equilibrium with a hot ambient medium of density $\rho_{\rm h}$.
We define the density contrast (equivalently, the overdensity) as $\chi \equiv \rho_{\rm c} / \rho_{\rm h}$ with typical values of $\chi \sim 10^2$ for a
warm neutral/ionized cloud $T \sim 10^4$\,K embedded in a hot soft X-ray phase $T \sim 10^6$\,K.
\footnote{A cold neutral cloud $T \sim 10^2$\,K exposed directly to a hot ambient medium $T \sim 10^6$\,K would result in a much larger $\chi \sim 10^4$. However, in practice a stable $10^4$\,K cocoon rapidly forms around the colder gas and hence still $\chi \sim 10^2$ \citep{farber2022survival}, although each phase samples a different portion of the cooling curve. 
Also note, the hot ambient phase is quite possibly hotter $T \sim 10^{7-8}$\,K.}
If the hot medium is a wind with relative velocity $v_{\rm rel}$, then the Kelvin-Helmholtz time is $t_{\rm KH} = \chi^{1/2} / (k v_{\rm rel})$ \citep[][]{Chandrasekhar1961}. 
Although small wavelengths grow fastest, $k^{-1} \sim R_{\rm cl}$ are the most destructive, so we define the destruction time as $t_{\mathrm{destroy}} \equiv \chi^{1/2} R_{\rm cl} / v_{\rm rel}$. 
For nonradiative strong shocks \citet{Klein1994} showed both the shock-crossing time and the Rayleigh-Taylor growth time are comparable to the Kelvin-Helmholtz time, which we generically call the cloud crushing time, $t_{\mathrm{cc}} \equiv \chi^{1/2}$ \rcl / \vrel.

Next, consider the equation of motion for a rigid cloud
\begin{math}
m_{\mathrm{cloud}}  \dot v_{\rm cloud}
= -\frac{1}{2} C_{\rm D} \rho_{\rm h} \vrel^2 A
\end{math}
where $C_{\rm D}$ is the drag coefficient $\sim 1$ and $A$ is the cross-sectional area of the cloud. Then it is easy to show that the acceleration time is $t_{\rm drag} = \chi \rcl / \vrel = \chi^{1/2} t_{\rm cc}$. Therefore, $t_{\rm drag} \gg t_{\rm cc}$.

Indeed, both early \citep{Cowie1977,nittmann1982dynamical,Stone1992,Klein1994} and recent studies of cold clouds in hot winds \citep{Scannapieco2015,bruggen2016launching,Girichidis2021} find cloud destruction, which is rather puzzling since fast moving cold gas is observed throughout the Universe \citep[see, e.g.,][]{Veilleux2020}.
One potential solution is the inclusion of magnetic fields as they purportedly suppress instabilities and thus mixing. 

For instance, \citet{Mac1994} studied strong MHD shocks impacting a $\chi \sim 10$ cloud and found putative cloud survival due to the magnetic field damping Kelvin-Helmholtz instability and braking vortexes (although they only ran their simulations to 3 $t_{\mathrm cc}$). 
\citet{shin2008magnetohydrodynamics} studied a variety of initial orientations (with respect to the shock front) and strengths of the magnetic field with high resolution $\sim 100$ cells per \rcl\ in three-dimensional supersonic (sonic Mach = 10) simulations. 
They found no difference even for strong fields during the first four $t_{\rm cc}$
but subsequently MHD reduces fragmentation and mixing. Weak $\beta \sim 10$ perpendicular fields produced significantly different morphological evolution of the clouds compared to parallel magnetic fields or hydrodynamic simulations.
This is in agreement with works finding magnetic fields suppress Richtmeyer-Meshkov \citep[][]{wheatley2005suppression}, Kelvin-Helmholtz \citep[][]{ryu2000magnetohydrodynamic} and Rayleigh-Taylor \citep[][]{stone2007nonlinear,stone2007magnetic} instabilities and corroborated by recent high-resolution studies \citep{Sparre2020}.\footnote{This in agreement with linear theory: \citet{Chandrasekhar1961} showed the Kelvin-Helmholtz instability is suppressed if the Alfvén speed exceeds the shear speed, or $\beta < 2 / M_{\rm s}^2$ for $\gamma = 5/3$ and the Rayleigh-Taylor instability is suppressed if $t_{\rm sc,alfven} < t_{\rm drag}$\ so $\beta < (2/\gamma)(\chi/M_{\rm s})^2$. Thus we should always find Rayleigh-Taylor suppressed (unless $\beta \gtrsim 10^4$) but only Kelvin-Helmholtz suppressed for dynamically important magnetic fields.} 
Interestingly, however, in three-dimensional magnetohydrodynamic simulations of mildly supersonic cold clouds in hot winds, \citet{gregori1999enhanced} found magnetic fields enhanced the growth rate of Rayleigh-Taylor instability by trapping of vortexes on surface deformations, causing more rapid cloud destruction. Albeit, their resolution was poor compared to e.g., \citet{shin2008magnetohydrodynamics}.

Another important effect of magnetic fields is the buildup of pressure upstream of the cloud and consequential faster acceleration \citep[][]{jones1996magnetohydrodynamics,fragile2005magnetohydrodynamic,Dursi2008,pfrommer2010detecting,McCourt2018}. Specifically, this process known as `magnetic draping' shortens the drag time to \citep{Dursi2008,McCourt2015}
\begin{equation}
   \frac{t^{\rm mhd}_{\rm drag}}{t^{\rm hydro}_{\mathrm{drag}}} = \left(1 + \frac{2}{\beta_{\mathrm{w}}\mathcal{M}^2}\right)^{-1}.
   \label{eq:tdrag_rat}
\end{equation}
where $\beta_{\rm w}$ is the ratio of the thermal to magnetic pressure in the wind. However, for survival, we require $t_{\rm cc} \sim t_{\rm drag}$, thus, the successful acceleration of cold gas with aid of magnetic fields only works for overdensities of
\begin{equation}
    \chi \lesssim 9 \left(\frac{\xi}{3}\right)^{2} \left(1 + \frac{2}{\beta_{\rm w}\mathcal{M}}\right)^2
    \label{eq:chi_crit}
\end{equation}
where we used $\xi \sim t_{\rm life} / t_{\rm cc}$ to parametrize the lifetime of the cloud. This yields $\chi\lesssim 32$ for a transonic $\mathcal{M}\sim 1.5$, $\beta_{\rm w}\sim 1$ wind. The above follows the argument of \citet{Gronke2020Cloudy}, who also checked Eq.~\ref{eq:chi_crit} using adiabatic cloud-crushing simulations with $\beta_{\rm w} = 1$.
 In conclusion, for most media of astrophysical interest where $\chi \gtrsim 100$, magnetic fields alone do not solve the entrainment problem, i.e., cold gas is destroyed prior to being accelerated. \\

However, another proposed solution to the `entrainment problem' is the inclusion of radiative cooling and indeed simulations with efficient radiative cooling generically show cold clouds can survive and even grow in hot winds if some cooling time is shorter than the destruction time. Studies on this regard include observational evidence from galactic fountains
\citep[]{Marinacci2010, Mandelker2020} and other exhaustive analysis of properties of the wind and clouds
\citep[]{Sparre2020, gronke2022survival, fielding2022structure, tan2022cloudy}. The study of appearance and structure of the entrained gas \citep[]{banda2016filament, Sparre2019, farber2022survival}, besides turbulence processes that induce such entrainment \citep[]{Kanjilal2021} have proven to be essential to the understanding of the problem. Much work has been put into hydrodynamical simulations 
\citep[]{Armillotta2017, Gronke2018, li2020simple}, and other more complex studies \citep[]{Abruzzo2021, li2020simple, abruzzo2022taming} with the purpose of reproducing the physics behind these observations.

Specifically, \citet{Gronke2018} have shown using hydrodynamical simulations that cold gas can survive when the following condition is satisfied:
\begin{equation}
\centering
\frac{t_{\rm cool, mix}}{t_{\rm cc}}  \lesssim 1
\label{eq:tcoolmix_tcc_crit}
\end{equation}
with $t_{\rm cool,mix}$ as the radiative cooling timescale for the mixing turbulent interface between multitemperature plasmas, namely $t_{\rm cool,mix}\equiv t_{\rm cool}(T_{\rm mix}, \rho_{\rm mix})$ with $T_{\rm cool,mix}=\sqrt{T_{\rm h}T_{\rm c}}$ and $\rho_{\rm mix}=\sqrt{\rho_{\rm h}\rho_{\rm c}}$ \citep{begelman1990turbulent,Hillier2019}. Other works \citep{Li2020,Sparre2020} introduce similar arguments but use different cooling and survival timescales \citep[see ][for a comparison]{Kanjilal2021}.

To connect Eq.~\eqref{eq:tcoolmix_tcc_crit} to observations, it is interesting to point out that $t_{\rm cc}\propto r_{\rm cl}$, whereas $t_{\rm cool}$ is merely a function of the properties of the gas. 
One can hence rewrite the survival criterion Eq.~\ref{eq:tcoolmix_tcc_crit} into a geometrical condition $r\gtrsim r_{\rm crit}$ \citep[cf.][]{Gronke2018,Gronke2020Cloudy} stating only clouds larger than a critical radius (dependent on the physical conditions) will survive the acceleration process.

While the effects of either radiative cooling or magnetic fields on ram pressure acceleration have been studied extensively, the combination of both has only been addressed in relatively few studies \citep{fragile2005magnetohydrodynamic,McCourt2015,Gronnow2018,cottle2020launching}. In particular, a systematic exploration of how magnetic fields affect the survival criterion Eq.~\eqref{eq:tcoolmix_tcc_crit} is outstanding. This is noteworthy as it has been argued that $r_{\rm crit}$ following from Eq.~\eqref{eq:tcoolmix_tcc_crit} applies to the survival of only relatively massive clouds \citep{Sparre2020,Xu2022ApJ...933..222X}, yet we know already that $r_{\rm crit}\rightarrow 0$ for $\chi\lesssim 30$, $\beta\sim 1$ and transonic winds \citep{Gronke2020Cloudy}.

In this paper, we aim to systematically clarify the interplay between radiative cooling and magnetic draping. The structure of this paper is as follows. In Sec. \ref{sec:methods} we divulge our numerical methods. We show our results in Sec. \ref{sec:results}, discuss in Sec. \ref{sec:discussion}, and conclude in Sec. \ref{sec:conclusions}.

\section{Methods}
\label{sec:methods}
\subsection{Numerical Methods}
We performed our simulations using the Eulerian grid code Athena 4.0 \citep{stone2008athena} on a three-dimensional regular Cartesian basis (with fixed grid spacing $d_{\mathrm{cell}}$), using an HLLD Riemann solver with third-order reconstruction and the constrained transport method to solve the compressible, inviscid, radiative fluid equations including a divergence-free magnetic field.

The \citet{townsend2009} algorithm is used for the integration of optically-thin radiative cooling included in our simulations, which requires a piecewise powerlaw solution for the cooling curve. Here we use the seven-piece power law fit to the \citet{Sutherland1993} cooling curve of \citet{McCourt2015} (also used by \citealt{Gronke2018}, \citealt{Gronke2020Cloudy}, and \citealt{farber2022survival}), assuming gas of solar metallicity. This cooling rate can be computed as:

\begin{equation}
    \Lambda (T) = c_k\left(\frac{T}{T_k}\right)^{\alpha_k}
\end{equation}
where $\Lambda (T)$ is the temperature-dependent cooling rate\footnote{Note that the \citet{townsend2009} method assumes cooling occurs isochorically during the simulation step $\Delta t$.}, $c_k$ is the power-law coefficient, $T_k$ is the lower bound of the temperature bin of the seven-piece power law fit and $\alpha_k$ is the power-law index.
Table \ref{tab:townsend} shows the temperature bins and coefficients employed.

\begin{table}
	\centering
	\caption{Piecewise power law fit to the cooling curve adopted in our simulations.}
	\label{tab:townsend}
	\begin{tabular}{lccr} 
		\hline
	    Temperature range & Coefficient (erg cm$^3$ s$^{-1}$ & Index\\
		\hline
        $8000 K \leq T < 10^4 K$ & $3 \times 10^{-26}$& 19.6 \\ 
        $10^4 K \leq T < 2 \times 10^4 K$ & $2.4 \times 10^{-24}$& 6\\
        $2 \times 10^4 K \leq T < 2 \times 10^5 K $& $1.5438 \times 10^{-24}$ & 0.6\\
        $2 \times 10^{5}K \leq T<1.5\times 10^{6}K$ & $6.6831 \times 10^{-22}$ & -1.7\\
        $1.5\times10^{6}K \leq T<8 \times10^{6}K$ &$2.7735 \times 10^{-23}$ & -0.5 \\
        $8\times10^{6}K\leq T<5.8\times10^{7}K$ & $1.1952 \times10^{-23}$& 0.22 \\
        $5.8\times10^{7} K \leq T$& $1.8421 \times 10^{-23}$ & 0.4 \\
        		\hline
	\end{tabular}
\end{table}

\subsection{Initial conditions} 
The computational setup is similar to past experiments by \cite{Gronke2020Cloudy}. 
A stationary, spherical cloud of radius $r_{\rm cl}$, temperature $T_{\rm cl} \sim 4\times 10^4$\,K and density $n_{\rm cl} \sim 0.1 \mathrm{cm}^3$ is embedded in pressure equilibrium with a hot wind with an overdensity $\chi = n_{\rm cl} / n_{\rm wind} = T_{\rm wind} / T_{\rm cl} = 100$ (and in a few cases $\chi = 10^3$). 
Additionally, we initialize a tangled $\sim$force-free magnetic field inside the cloud following \citet{McCourt2015} \& \citet{Gronke2020Cloudy} with a magnetic coherence length $r_{\rm cl}/10$ and strength $\beta_{\rm cl} = \frac{P}{P_{\rm B}}$, where $P$ is the thermal pressure and $P_{\rm B}$ is the magnetic pressure. 

A numerical resolution of $r_{\rm cl}/d_{\rm cell}$ = 16 is kept constant throughout the domain dimensions. This resolution is proven to converge for cold-gas mass evolution in previous works \citep{Gronke2020Cloudy} (see Appendix \S~\ref{sec:numericalconv} for a convergence study).
The hot wind is travelling at a transonic Mach number $\mathcal{M} = v_{\rm wind}/c_{\rm s,h}$ of 1.5.
Contrasting to past experiments, no cooling ceiling is imposed (whereas \citealt{Gronke2020Cloudy} turn off cooling in most of their simulations above 0.6 $T_{\rm wind}$; we confirmed the wind negligibly cools throughout the duration of our simulations).

The simulation domain consists of a three-dimensional rectangular box of fiducial size 64 \rcl\ x (12 \rcl)$^2$.  A zero-gradient outflow boundary condition is applied with the exception of the -$x$ face which applies the wind conditions. There is initially no magnetic field in the wind, but the inflow includes a magnetic field perpendicular to the wind axis with a strength that we fix to the plasma beta initialized in the cloud $\beta_{\rm wind} = \beta_{\rm cl}$. The evolution of cold gas clumps was demonstrated to be majorly determined by the magnetic strength of the incoming wind plasma by \citep{Gronke2020Cloudy}. We choose $\beta_{\rm cl} $ and $\beta_{\rm wind}$ to be equal for simplicity. For the low $\beta$ simulations the size of the box perpendicular to the wind axis was enlarged to 24 \rcl\ per dimension to reduce possible gas outflows orthogonal to the wind-axis from the simulated domain. 

\begin{table}
	\centering
	\caption{Simulation parameters. The final status is determined from the change in dense gas mass ($n > n_{\rm cl}/3$). For all simulations we use $r_{\rm cl}/d_{\rm cell}$ = 16, Mach number ($\mathcal{M}$) = 1.5, $\chi = 100$ and $T_{\rm cl} \sim 10^4\,K$ so $T_{\rm wind} \sim 10^6$\,K. Thereafter, runs share a fiducial $t_{\rm cool,mix}$ = 0.06\, Myrs and $t_{\rm cool,wind}$ = 10 Myrs}
	\label{tab:example_table}
	\begin{tabular}{lcccccr} 
		\hline
		$\beta$ & $t_{\rm cool,mix}/t_{\rm cc}$ & $r_{\rm cl}$  (pc) & $t_{\rm cc}$ (Myr) & status\\
		\hline
		1 & 0.1 & $7$  & $0.6$ & survived\\
		1 & 3 & $0.2$ & 0.02  & survived\\
  		1 & 30 & $0.02$ & 0.002  & survived\\
		1 & 100 & $0.006$ & $0.0006$  & survived\\
		1 & 300 & $0.002$ & 0.0002  & destroyed\\
		1 & 1500 & $0.0005$ & $4 \cdot 10^{-5}$  & destroyed\\
		10 & 0.1 & $7$ & 0.6  & survived\\
		10 & 1 & $0.7$ & 0.006  & survived\\
		10 & 3 & $0.2$ & 0.02  & survived\\
		10 & 20 & $0.03$ & 0.003  & borderline\\
        10 & 100 & $0.006$ & $0.0006$ & destroyed\\
        $10^{2}$ & 0.1 & $7$  & $0.6$ & survived\\
        $10^{2}$ & 1 & $0.7$ & 0.006   & survived\\
        $10^{2}$ & 3 & $0.2$ & 0.02   & survived\\
        $10^{2}$ & 10 & $0.07$ & 0.006  & survived\\
        $10^{2}$ & 30 & $0.02$ & 0.002 & borderline\\
        $10^{2}$ & 100 & $0.006$ & $0.0006$  & destroyed\\
        $10^{3}$ & 0.1 & $7$  & $0.6$ & survived\\
        $10^{3}$ & 3 & $0.2$ & 0.02  & survived\\
        $10^{3}$ & 30 & $0.02$ & 0.002  & destroyed\\
        $10^{3}$ & 100 & $0.006$ & $0.0006$   & destroyed\\
        $10^{4}$ & 0.1 & $7$  & $0.6$ & survived\\
        $10^{4}$ & 1 & $0.7$ & 0.006   & survived\\
        $10^{4}$ & 3 & $0.2$ & 0.02   & destroyed\\
        $10^{4}$ & 30 & $0.02$ & 0.002  & destroyed\\
        $10^{20}$ & 0.1 & $7$  & $0.6$  & survived\\
        $10^{20}$ & 3 & $0.2$ & 0.02  & destroyed\\
        $10^{20}$ & 100 & $0.006$ & $0.0006$ & destroyed\\
        $10^{20}$ & 1000 & $6 \cdot 10^{-4}$ & $6 \cdot 10^{-5}$  & destroyed\\
        $10^{20}$ & $10^{4}$ & $7 \cdot 10^{-5}$ & $6 \cdot 10^{-6}$  & destroyed\\
		\hline
	\end{tabular}
\end{table}

\begin{figure*}
    \includegraphics[width=\textwidth]
    {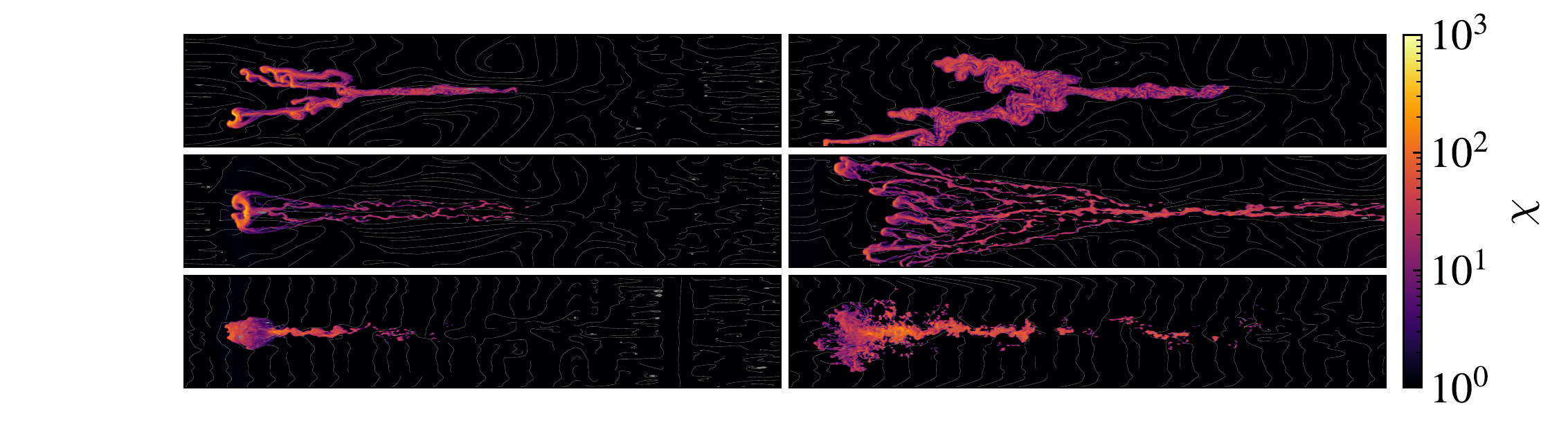}
        \caption[width=\textwidth]{Morphology plots for simulations of \tratio$=0.1$  and $\beta=1,100$ and $10^{20}$ from top to bottom, respectively. Each figure represents a projection through the z-axis of overdensity $\chi$ at times $t = 5, 10 \ t_{\rm cc}$ from left to right. Grey lines represent the contour lines for magnetic field lines in our runs. Earlier times exhibit higher compression of the lines near the cloud interface, shaping and draping overdense material under the influence of the wind.}
    \label{fig:multislice}
\end{figure*}

\subsection{Cloud tracking system}
\label{subsec:scalar}
To minimise the amount of cold cloud material flowing outside the domain, we use a cloud-tracking system putatively described in \cite{Gronke2018} (and in more depth in other studies e.g., \citealp{dutta2019modelling}).
We initially assign a passive, Lagrangian scalar concentration $C$=1 to the cold plasma \citep{Xu1995}.
This variable is equally subject to the MHD equations via the influence of the MHD equations on the background gas velocity it advects with. From the mass continuity equation, it is possible to deduce the average cloud speed $u_{\rm cl}$:

\begin{equation}
\langle u_{\rm cl} \rangle= \frac{\int_{x_{\rm min}}^{x_{\rm cl,0}}u_x C_{\rm cl}\rho dV }{\int_{x_{\rm min}}^{x_{\rm cl,0}}C_{\rm cl}\rho dV}
\end{equation}
$x$ is the coordinate basis for the wind speed direction, $x_{\rm cl,0}$ is the projected initial cloud centre position in the $x$-axis, $x_{\rm min}$ is the minimum $x$ value of the domain, $u_{x}$ is the velocity in the $x$-direction, $C_{\rm cl}$ is the initial maximum concentration of gas divided by three and $dV$ is the infinitesimal (cell) volume. 

Similar to how it is described in previous work \citep{Scannapieco2015,bruggen2016launching,McCourt2015,farber2022survival}, this tracking method allows for a significant reduction in the box size required to contain the cloud material while reducing advection errors \citep{robertson2010}.

\section{Results}
\label{sec:results}
In this section we investigate the interplay between radiative cooling and magnetic fields (parametrized via $t_{\rm cool,mix}/t_{\rm cc}$ and $\beta$, respectively). We focus in particular on their impact on cloud gas survival.

\subsection{Morphological Evolution}
We begin by selecting simulations at fixed $t_{\rm cool,mix}/t_{\rm cc} = 0.1$ to explore the dependence of the morphological evolution of clouds on relative magnetic field strength. In Figure \ref{fig:multislice} we display projections of the overdensity $\chi$ for $\beta = 1, 100 \ \text{and} \ 10^{20}$ from top-bottom with dark regions of low overdensity and bright regions of high overdensity. We overplot magnetic field vectors as grey curves to explore the dependence of cloud morphology on relative magnetic field strength. The columns show different times in the evolution: $t = 5 , 10  \ t_{\rm cc}$ from left to right.

 The left column shows snapshots at time $t = 5 t_{\rm cc}$. The top cloud corresponding to $\beta = 1$ has a filamentary morphology with fairly weak magnetic fields outside the cloud but strong magnetic fields inside the filamentary cloud material. The second top row for $\beta = 100$ shows the cloud has broken into several `head-tail' structures which appear to be more `cored' or chunky in the tail material; that is, the tail is not continuous cloud material. The magnetic field appears stronger over a wider area, corresponding to the greater lateral distribution of cloud material. In all cases of $\beta$ the magnetic field appears to be draping the cloud material.

The right column shows  time $t = 10 t_{\rm cc}$. While the top row for $\beta = 1$ again shows a single filament of cloud material, the second row from the top $\beta = 100$ has a much broader lateral distribution of thin filaments looking like a spider's web in morphology. For all three scenarios, Rayleigh-Taylor 3D instabilities are not suppressed under the influence of magnetic fields. This is in accordance to previous studies on magnetic RT instabilities \citep{stone2007magnetic}, suggesting that the nonlinear regime can even be amplified in the presence of magnetised gas.
The bottom row still has strings of isolated cores of cloud material in contrast to the continuous filament in the top row. While all of these clouds survive, the rather different morphological evolution from the top row to the bottom rows suggests that cloud survival may be impacted by the morphological dependence on $\beta$.

\subsection{Cloud Mass Evolution}

\begin{figure}
        \includegraphics[width=\columnwidth]{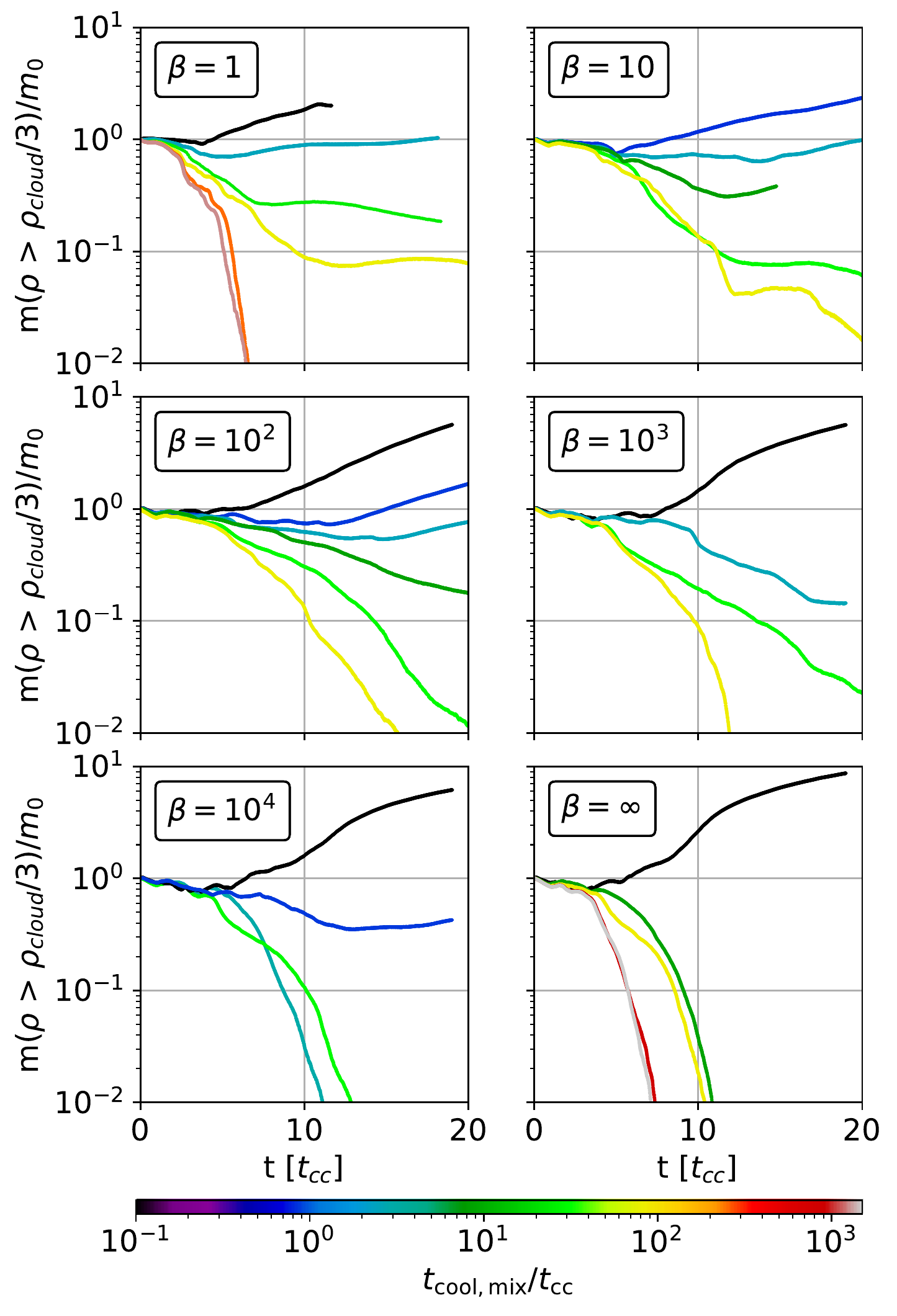}
    \caption{Cold gas mass evolution of clouds exposed to a $\mathcal{M}\sim 1.5$ wind and overdensity of $\chi \sim 100$ with different plasma $\beta$ (from $\beta \sim 1$ in the upper left to $\beta=\infty$ in the lower right panel) and varying cooling strength (indicated by the line color).}
    \label{fig:mass_evolution_trial}
\end{figure}

To further investigate the dependence of cloud survival on the interplay between radiative cooling ($t_{\rm cool,mix}/t_{\rm cc}$) and magnetic fields, we next explore the mass evolution of the full suite of simulations covering the \tratio-$\beta$ parameter range.
Figure~\ref{fig:mass_evolution_trial} shows the cloud gas, specifically $m(\rho > \rho_{\rm cloud}/3$), against time in units of $t_{\rm cc}$.

Each panel in Fig.~\ref{fig:mass_evolution_trial} represents  a different magnetic field strength with $\beta$=1, 10, 100, 1000, $10^4$, and $10^{20}$ ordered from top-left to bottom-right, respectively. 
In each panel we plotted the simulations performed at fixed $\beta$ and varying \tratio\ (denoted via the color coding) to specifically determine the influence of $\beta$ on the critical \tratio\ for survival of cloud material. The limiting cooling ratio for survival is expected to evolve with magnetic $\beta$. We selected  \tratio\ values that are inferred to lie closer to the survival limit for each $\beta$ case from neighbouring $\beta$ simulations, as opposed to performing them systematically for every possible \tratio, which would outstandingly increase the computational cost of finding the limiting criterion for each $\beta$.

The top-left panel of Fig.~\ref{fig:mass_evolution_trial} displays simulations performed at fixed plasma beta of $\beta = 1$ 
with five \tratio\ values of 0.1, 3, 100, 300 and 1500.
Interestingly, we find clouds survive (are destroyed) for \tratio\ $\lesssim$ 100 ($\gtrsim$ 100), whereas hydrodynamic simulations find destruction for \tratio\ $\gtrsim$ 1 \citep[][and also the lower right panel of Fig.~\ref{fig:mass_evolution_trial}]{Gronke2018}.
Clouds with inefficient cooling (\tratio\ $\gtrsim$ 100) display a rapid destruction. Clouds with efficient cooling, particularly, our run for $t_{\rm cool,mix}/t_{\rm cc} \sim 0.1$ exhibits an initial period of mass loss followed by 
mass growth. 
In the case of $t_{\rm cool,mix}/t_{\rm cc} \sim 3$, similar characteristics are observed, with the exception of apparent episodic saturation in growth of the cloud. 
These ‘bumps’ are not observed in previous works \citep{Gronke2018} wherein once any evolving cloud begins growing in mass, the cloud continues growing in mass monotonically. These saturation interludes might be related to magnetic fields impacting the pulsations observed in earlier work. 
This may also be why \tratio\ = 100 saturates at a reduced mass, as mass growth stagnates if there are no pulsations driving mixing. While oscillations in the mass loss rate can also be observed for dying clouds, the clouds are still rapidly destroyed.

Results for $\beta = 10$ are shown in the top right panel of Fig.~\ref{fig:mass_evolution_trial}. Multiple runs were performed, varying \tratio\ from 0.1 to 100. Growth in cloud material is observed for simulations with timescale ratios $\lesssim$ 10, contrasting with destroyed clouds for $t_{\rm cool,mix}/t_{\rm cc} = 100$, and the \tratio\ = 20 case was restarted several times but remains difficult to say if it will eventually lose all its dense gas or instead survive.
For the \tratio\ = 3 case, surviving gas experiences a critical point at $t\sim 13 t_{\rm cc}$ in which it transitions from mass loss to growth. 
The aforementioned oscillatory behaviour in the mass curves can here be clearly noticed for $t_{\rm cool,mix}/t_{\rm cc} = 100$, with ridges momentarily changing the tendency from mass loss to growth, yet the cloud still is ultimately destroyed.

The left-hand-side, second row panel shows the status for $\beta = 100$ with \tratio\ of 0.1, 1, 3, and 10 surviving, $t_{\rm cool,mix}/t_{\rm cc}$ $\sim$100 destroyed, and \tratio\ = 30 which is again unclear despite several restarts.

The central right panel of Fig.~\ref{fig:mass_evolution_trial} shows the runs with $\beta = 1000$ for $t_{\rm cool,mix}/t_{\rm cc}$ values of 0.1 and 100. 
Here, the only surviving clouds are labelled as dark blue ($\beta = 0.1$)
and sky-blue ($t_{\rm cool,mix}/t_{\rm cc} = 3$), 
while the dying clouds, yellow, corresponding to 
$t_{\rm cool,mix}/t_{\rm cc} = 100$ and light-green (\tratio\ = 30). 

The bottom-left panel of Fig.~\ref{fig:mass_evolution_trial}, representing $\beta = 10^4$, consists of runs with $t_{\rm cool,mix}/t_{\rm cc} = 0.1, 1, 3$ and 30.
The first two evidently survive, whereas the last two $t_{\rm cool,mix}/t_{\rm cc} = 3$ and 30 are completely ablated by the hot gas. 
Sharper oscillatory behaviour is observed for this order of plasma beta values.
Hence, for $\beta \sim 10^4$ we roughly recover the hydrodynamic survival criterion \tratio\ $\lesssim 1$ \citep{Gronke2018}.

Lastly, the right-hand-side bottom panel of figure \ref{fig:mass_evolution_trial} displays simulations for $\beta = 10^{20}$, serving for comparative purposes to hydrodynamic runs from \citet{Gronke2018}. The smooth dying nature of cold gas for $t_{\rm cool,mix}/t_{\rm cc}= 3, 100, 1000$, and $10^4$ (green, yellow, red and silver) highly contrasts with the jagged mass increase of $t_{\rm cool,mix}/t_{\rm cc} = 0.1$ (indigo), similar to the MHD runs. The destruction timescale for extremely weak cooling clouds are of similar duration, as are, although somewhat longer, for weak cooling \tratio\ = 3 and 100.
\\

Figure \ref{fig:mass_evolution_trial} demonstrates that the inclusion of magnetic fields increase the critical $t_{\rm cool,mix }/ t_{\rm cc}$ value for cold gas survival dramatically at low $\beta$ and by a factor of a few even for as high $\beta$ as $\sim$10$^3$.
However, cloud destruction via Rayleigh-Taylor and Kelvin-Helmholtz instability cannot be completely suppressed, as we find destruction in all cases of $\beta$ we simulated for sufficiently large \tratio\ (i.e., going towards the adiabatic limit). Note that \citet{Gronke2020Cloudy} found even adiabatic simulations of clouds survive, but only for low overdensities $\chi \lesssim 30$ (cf. Eq. \ref{eq:chi_crit}), whereas we simulate high overdensities $\chi = 100$.
Although cold gas can die even at plasma beta $\beta$ values of 1, the critical $t_{\rm cool,mix}/t_{\rm cc}$ increases with decreasing $\beta$ suggesting a strong connection for survival between radiative cooling and magnetic fields, which we make more clear next.

\subsection{Magnetic Fields Enhance Survival of Radiative Clouds}
\begin{figure}
\includegraphics[width=\columnwidth]{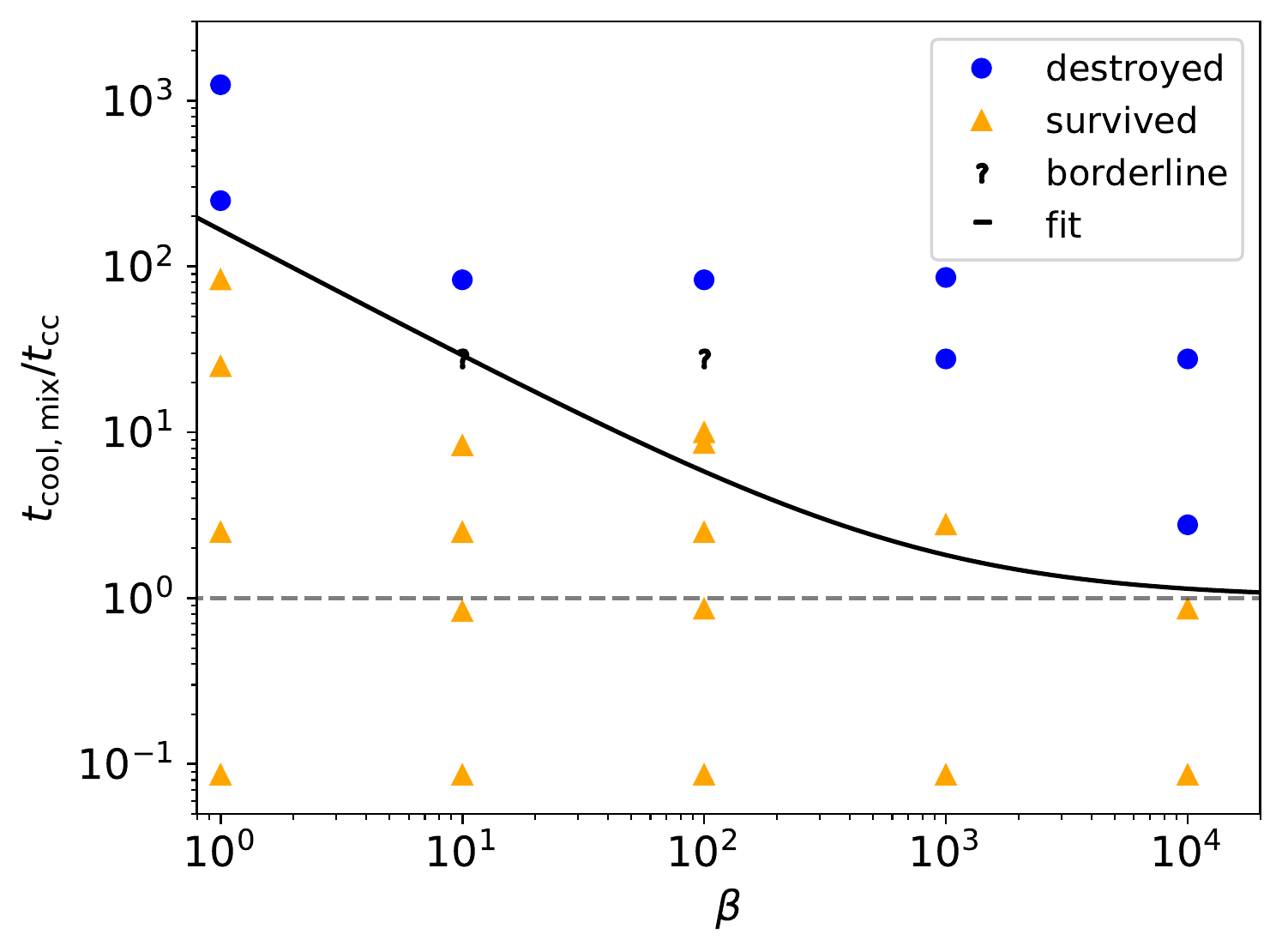}
    \caption{$t_{\rm cool,mix}/t_{\rm cc}$ against plasma $\beta$. The final state of the clouds are represented by blue dots in the case of survived, orange triangles for dissipated gas, black question marks for borderline cases, and we show the best fit powerlaw $t_{\rm cool,mix}/t_{\rm cc} = 1.0 + 200 \beta^{-0.7}$ as a black curve.}
    \label{fig:trat_trial}
\end{figure}

Figure \ref{fig:trat_trial} displays an overview of the \tratio-$\beta$ parameter space with the symbols indicating cloud survival or destruction clearly demonstrating that magnetic fields allow survival for less efficient radiative cooling than required in the hydrodynamic case \citep{Gronke2018}.
In Fig.~\ref{fig:trat_trial}, clouds that are destroyed are represented by blue dots whereas for the case of survived, these are shown as orange squares. Information in the axis is in ordered, logarithmic scale, with $\beta$ varying from 1 to 
$10^4$ and a range of $t_{\rm cool,mix}/t_{\rm cc}$ scaling from 0.1 to 1500. Simulations with plasma beta of $10^{20}$, a numerical approximation to purely hydrodynamical scenarios, are excluded for ease of presentation.

The curve delimiting the approximate boundary between destroyed and survived clouds $t_{\rm cool,mix}/t_{\rm cc} = 1.0 + 200 \beta^{-0.7}$ represents how the survival timescales of cold gas can vary as a function of plasma beta, $\beta$. Reading figure \ref{fig:trat_trial} from right to left in the plasma beta axis, i.e. coming down to lower betas from simulations nearer an ideal hydrodynamical scenario of $\beta = \infty$, reveals an upwards shift in the peaking timescales survival ratio of the clouds, becoming higher when approaching $\beta = 1$.

For weak magnetic fields with $\beta = 10^{4}$ we recover the hydrodynamical results with a critical $t_{\rm cool,mix}/t_{\rm cc}$ value of 1. 
Taking some distance from the hydrodynamic limit, for $\beta = 10^{3}$, the critical time ratio for survival is $t_{\rm cool,mix}/t_{\rm cc} \sim 3$.  
Remarkably, for $\beta=1$, the critical \tratio allowing cold gas survival is approximately 100 times that predicted in the absence of magnetism, $t_{\rm cool,mix}/t_{\rm cc} \approx 1$ \citep[][]{Gronke2018}. This case might be most relevant in the ISM where equipartition in thermal and magnetic energy densities is expected \citep{Draine2011}. We did not perform simulations below $\beta$ unity as these values are less relevant in galactic astrophysics (and numerically more challenging to evolve stably). In an attempt to better understand why magnetic fields produce such substantial changes in the survival criterion, we next turn to the entrainment time of these gas clouds.

\subsection{Entrainment Time for Magnetized Clouds}
\label{sec:results_entrainment}
\begin{figure}
        \includegraphics[width=\columnwidth]{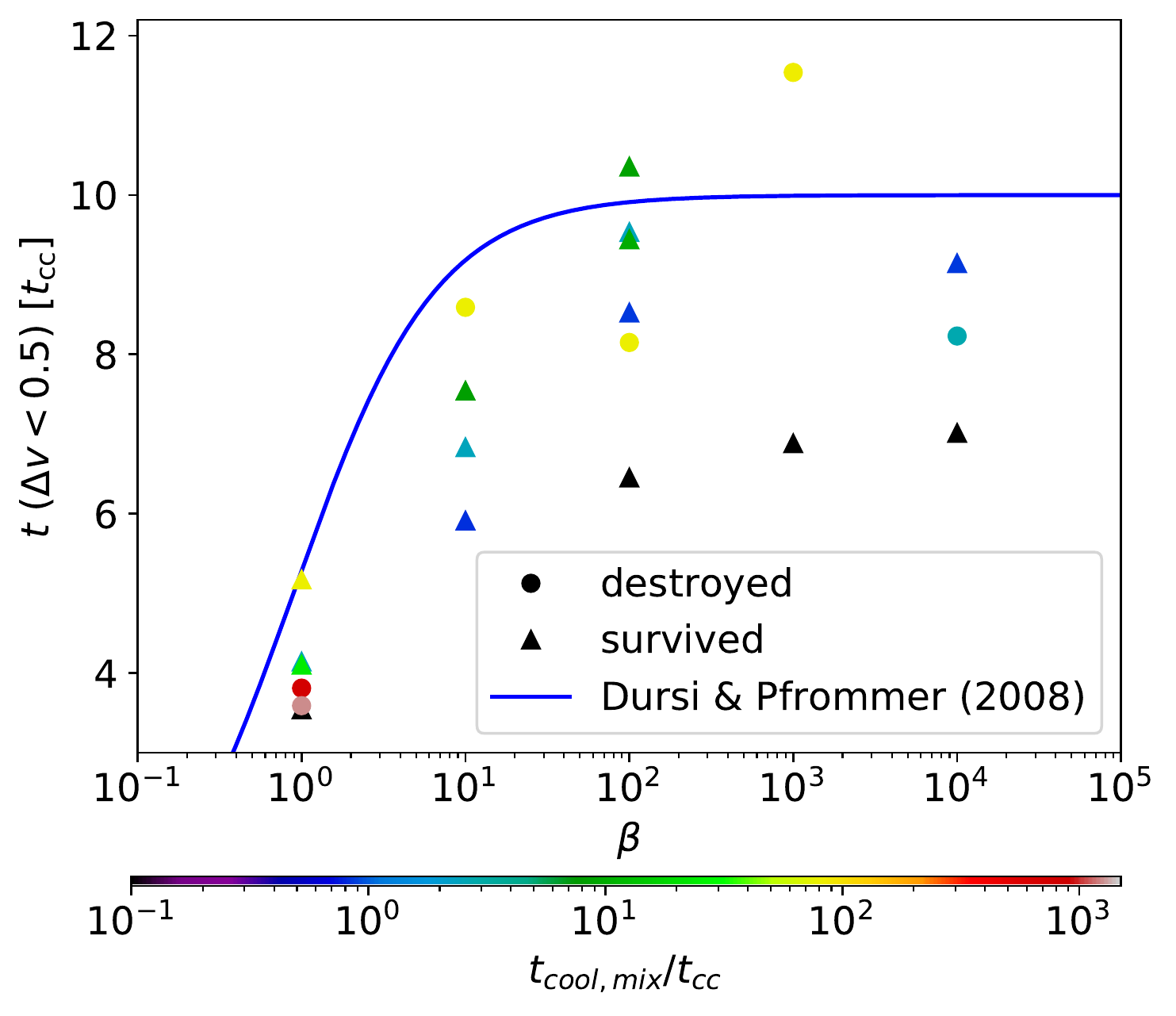}
    \caption{Time for the cold gas to reach $v = v_{\rm wind}/2$ vs $\beta$. Analogously to Figure \ref{fig:mass_evolution_trial}, the value of \tratio is indicated by the colours of the symbols, and similar to Figure \ref{fig:trat_trial}, we indicate clouds that survive as triangles and clouds that are destroyed as circles. The solid line shows the analytical estimate Eq.~\ref{eq:tdrag_rat}.}
    \label{fig:scatter}
\end{figure}

An overview of the clouds' velocity shear is represented in Figure \ref{fig:scatter}. 
That is, we plot the time required for the cloud to reach half the wind velocity as a function of $\beta$. 
As previously, we plot surviving clouds as triangles and and destroyed clouds as circles, with colors representing the $t_{\rm cool,mix}/t_{\rm cc}$ ratio.

Figure~\ref{fig:scatter} shows that the acceleration process depends strongly on the magnetization of the gas. Specifically, the entrainment time of the dense gas reaches its minimum for $\beta = 1$, followed by a saturation for $\beta \gtrsim 10$ where $t_{\rm drag} \sim \chi^{1/2} t_{\rm cc}$ as expected from analytical theory \citep{Dursi2008}.

While the trend is not monotonic, we do observe the fastest cooling runs show the most rapid entrainment, consistent with expectations since mass (and momentum) transfer $\dot{m} \propto t_{\rm cool}^{-1/4}$ \citep{Gronke2020Cloudy,tan2021radiative}. Note, however, that this dependence is not very strong -- and so the decrement in entrainment time is also not very large (i.e., two orders of magnitude change in cooling time only amounts to a factor of two reduction in the entrainment time).

Comparing the inefficient cooling cases, we find agreement with theoretical predictions from magnetic draping \citep[cf. Eq.~\ref{eq:tdrag_rat}][]{Dursi2008,pfrommer2010detecting,McCourt2015,Sparre2020}. The overall tendency matches the analytical prediction from equation ~\ref{eq:tdrag_rat}, only displaying slight deviations of a few tcc due to the varying cooling efficiencies which, as aforementioned, has a minor impact on the entrainment time (see Appendix \S~\ref{app:entrainmentTime} for an extended analysis). Yet the factor of two decrease in entrainment time does not evidently explain the $\sim$100 fold increase in survival for $\beta = 1$ simulations and particularly not the $\sim$10 fold increase in survival for $\beta = 100$ which is not appreciably accelerated faster than the hydrodynamic limit. To further explore what may bring about the increased survival for magnetized clouds we next investigate the mass accretion rate dependence on $\beta$ to investigate possible suppression of mixing by magnetic fields.

\subsection{Mass Accretion Rate}

\begin{figure}

        \includegraphics[width=\columnwidth]{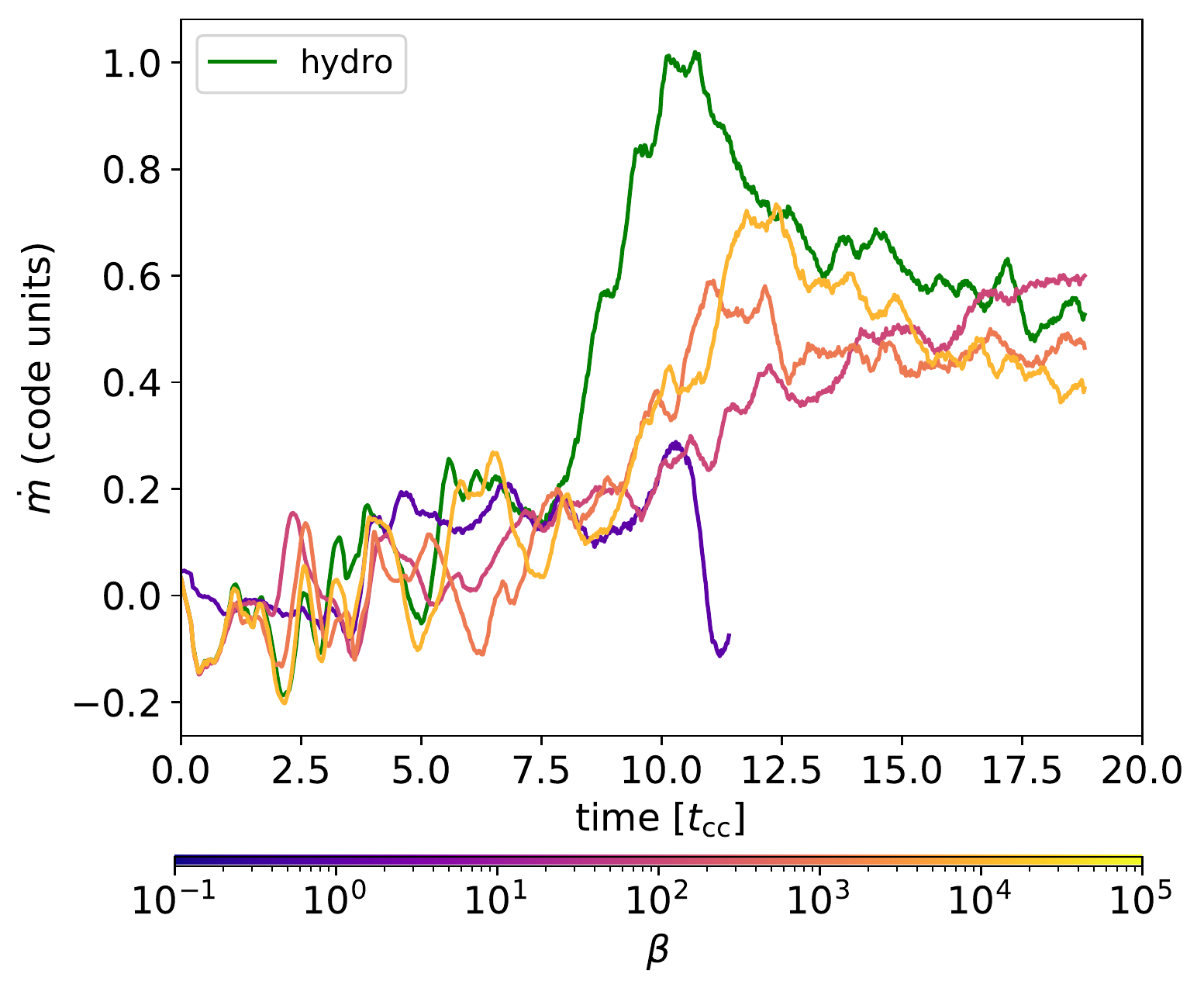}
    \caption{Mass growth rate of clouds (defined as material with $\rho > \rho_{\rm cl}/3$) as a function of time for a fixed $t_{\rm cool,mix}/t_{\rm cc}=0.1$ and varying $\beta$ (a hydro run is the green line for comparison).}

    \label{fig:mdot}
\end{figure}

In Figure \ref{fig:mdot}, we show the mass accretion rate against the time elapsed in units of $t_{\rm cc}$. We aim to explore how varying $\beta$ conditions may modify gas accretion from hydrodynamic instabilities, for which a fixed $t_{\rm cool,mix}/t_{\rm cc}$ ratio of 0.1, observed to survive regardless of plasma beta, is chosen as value of reference. We indicate $\beta$ by colours as previously and include a hydro run as a green curve for comparison purposes.

The overall picture reveals that $\dot m$ is fairly insensitive to $\beta$ with the mass growth in the hydrodynamical simulation maximally a factor of $\sim 10$ larger than in the low $\beta$ run and a factor of a few larger than high $\beta$ runs, but most of the time only tens of percent difference. This is somewhat surprising as linear analysis and shearing box simulations show that mixing is suppressed (after magnetic fields amplify and possibly the turbulent velocity has diminished) even with initially weak magnetic fields \citep{Ji2018,gronnow2022role} -- but consistent with a previous study \citep{Gronke2020Cloudy}.

\section{Discussion.}
\label{sec:discussion}

\subsection{The survival of cold gas in magnetized galactic winds}
\label{sec:disc_survival}
Our findings show that the combination of radiative cooling and magnetic fields allows $T\sim 10^4\,$K cold gas to survive in a hot ($T\sim 10^6\,$K) wind if
\begin{equation}
    \tratio \lesssim \frac{200}{\beta^{0.7}}\ + 1.
    \label{eq:tratio_fit}
\end{equation}
This implies that the physical parameter space for survival is increased drastically which is clear when rewriting Eq.~\eqref{eq:tratio_fit} to a geometrical criterion yielding
\begin{equation}
    r_{\rm cl} \gtrsim R_{\rm crit,cl} \sim 2\textrm{\,pc} \frac{T_{cl,4}^{5/2} \mathcal{M}}{P_3 \Lambda_{\rm mix,-21.4}} \textrm{min}\left(1, \frac{\beta^{0.7}}{200}\right). 
    \label{eq:rcrit_fit}
\end{equation}
Here, $T_{\rm cl,4}\equiv T/(10^4\,{\rm K})$ is the cloud temperature in units of $10^4$\,K, $\mathcal{M}$ is the Mach number, $P_3 \equiv P/(10^3\,{\rm cm}^{-3}\,\rm{ K}$) is the pressure (per Boltzmann constant) in units of $10^3$\,cm$^{-3}$\,K, $\Lambda_{\rm mix,-21.4}$ is the cooling rate in units of $10^{-21.4}$\,erg\,cm$^3$\,s$^{-1}$, and the last term includes the effects of magnetic fields, which impact the survival when $\beta \lesssim 10^4$.

This strong effect on the survival criterion might be surprising given our other results:
\begin{enumerate}
    \item \label{item:entrainment} While magnetic fields aid in cloud entrainment this effect is only $\mathcal{O}(1)$ (cf. Fig.~\ref{fig:trat_trial} and Eq.~\ref{eq:tdrag_rat}), i.e., at first sight too weak to explain the $\sim 2$ order of magnitude change in survival criterion for $\beta \sim 1$.

    \item Similarly, one expects magnetic fields to suppress turbulence and hence inhibit mixing. However, contrasting the evolution of the turbulent velocity, measured as the velocity orthogonal to the wind axis, for various selections of data ($\rho > \rho_{\rm cl}/3$, $T < 2 T_{\rm fl}$, and $T < T_{\rm mix}$) revealed no significant differences between simulations with magnetic fields compared to hydrodynamic cases (see Appendix \S~\ref{app:turbulence}). 

    \item We also checked if other effects such as compression due to magnetic fields and hence a lowered $t_{\rm cool}$ play a role but only found a small ($\lesssim$ factor of a few) change for relatively short durations ($\sim$1 - 2\,$t_{\rm cc}$). Since we expect the mass growth of the cold medium $\dot m\propto t_{\rm cool}^{-1/4}$ \citep{Gronke2020Cloudy,tan2021radiative,Fielding2020} or at most $\dot m\propto t_{\rm cool}^{-1/2}$ \citep{Ji2018,tan2021radiative}, we consider this effect to be negligible. 

    \item \label{item:mass} We do not find large changes in the mass evolution of cloud material when including magnetic fields. This is suggested by Fig.~\ref{fig:mass_evolution_trial} when comparing the mass evolution for runs with weak cooling. For instance, independent of $\beta$, the $\tratio\sim 10^3$ simulations lose all cloud mass by $t\sim 8t_{\rm cc}$. This is also consistent with the fact that we observe similar mass growth (and thus mixing) rate for the low-$\beta$ runs in Fig.~\ref{fig:mdot} for most of the time evolution, amounting to at most a factor of 2 suppression in mass growth when including magnetic fields compared to the hydrodynamic case in Fig.~\ref{fig:mass_evolution_trial}. However, note that these are limiting cases -- either specific times or very weak cooling -- and thus less relevant to the survival threshold which is (per definition) a limiting case.
\end{enumerate}

In summary, we have found two effects clearly altered by the presence of magnetic fields: the acceleration process and the mass evolution (discussed in \ref{item:entrainment} and \ref{item:mass} above, respectively). However, both of these effects are seem rather weak and not sufficient to explain the large change in the survival threshold. Below, we will analyze both effects in more detail.\\

\begin{figure}
    \centering
    \includegraphics[width=0.9\columnwidth]{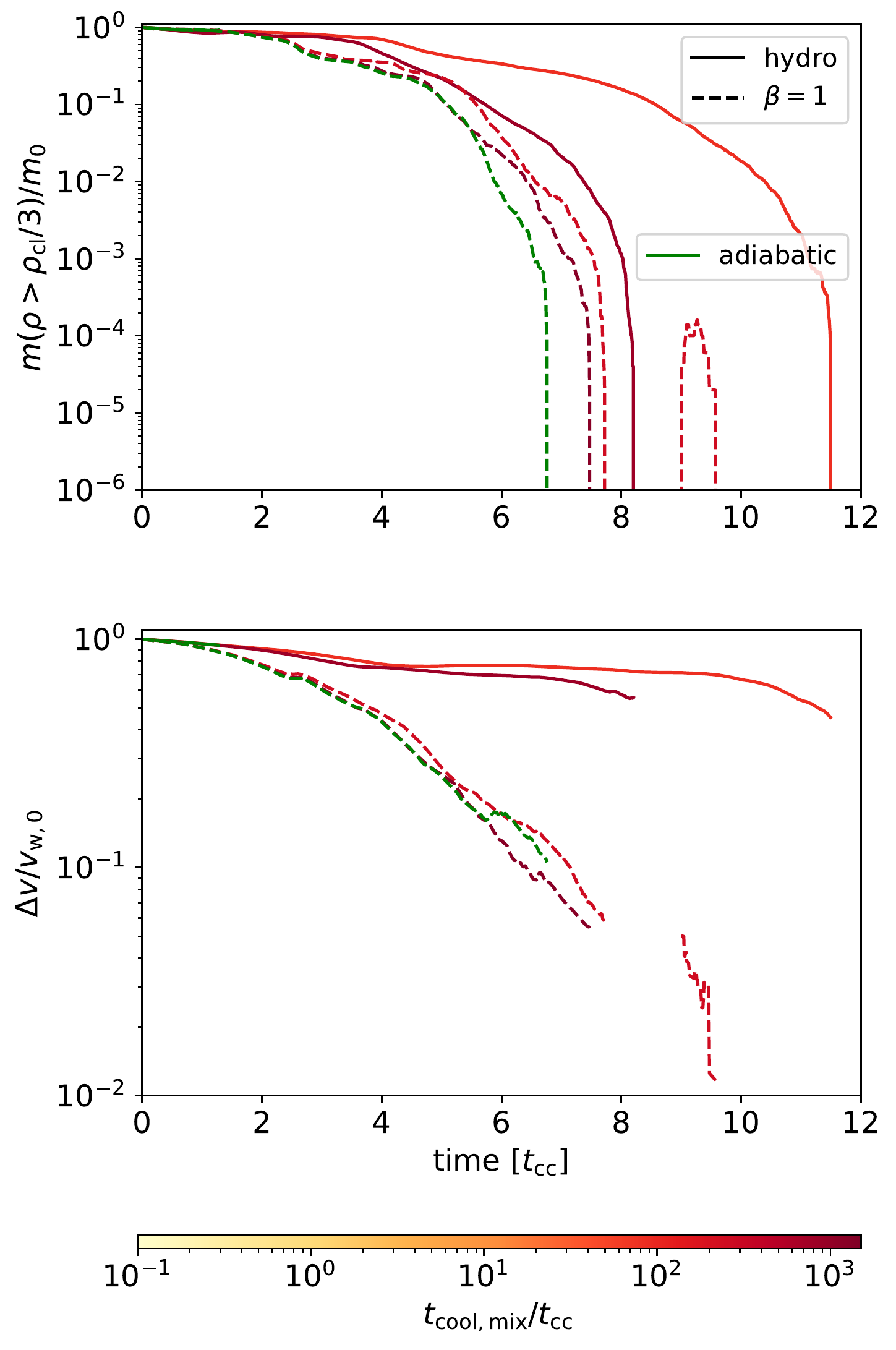}
    \caption{Time evolution of the mass (top panel) and relative velocity between the wind and cloud (bottom panel) for clouds which are destroyed in hydrodynamic runs and $\beta = 1$ as solid and dashed curves respectively, as a function of \tratio\ (note that we mark the adiabatic runs as green and the we changed colormaps since otherwise the adiabatic run may lead to confusion).
    }
    \label{fig:disc:weakCooling}
\end{figure}

First, we focus on these effects in the weak / no cooling limit. Fig.~\ref{fig:disc:weakCooling} shows the mass and velocity evolution for very weak and no cooling cases contrasting the pure hydrodynamical and $\beta\sim 1$ runs. Clearly, magnetic fields do speed up entrainment (cf. \S\ref{sec:results_entrainment}) but do not prolong the destruction process\footnote{Interestingly, the adiabatic $\beta\sim 1$ shown in Fig.~\ref{fig:disc:weakCooling} even shows the fastest destruction which likely is due to the faster acceleration and consequently enhanced Rayleigh-Taylor instability.}. Only with $\tratio\sim 100$ (red curve in Fig.~\ref{fig:disc:weakCooling}) the mass evolution is clearly altered. Notice that discontinuities in entrainment curves are a direct consequence of our cold gas criterion: for some time intervals, overdense gas can become diffuse enough not to satisfy the $m(\rho > \rho_{\rm cl}/3)$ condition. This appears as gaps in the velocity shear tendency. As it progressively accretes gas from the medium, it can make the cut and be displayed in the figure again.

\begin{figure}
        \includegraphics[width=\columnwidth]{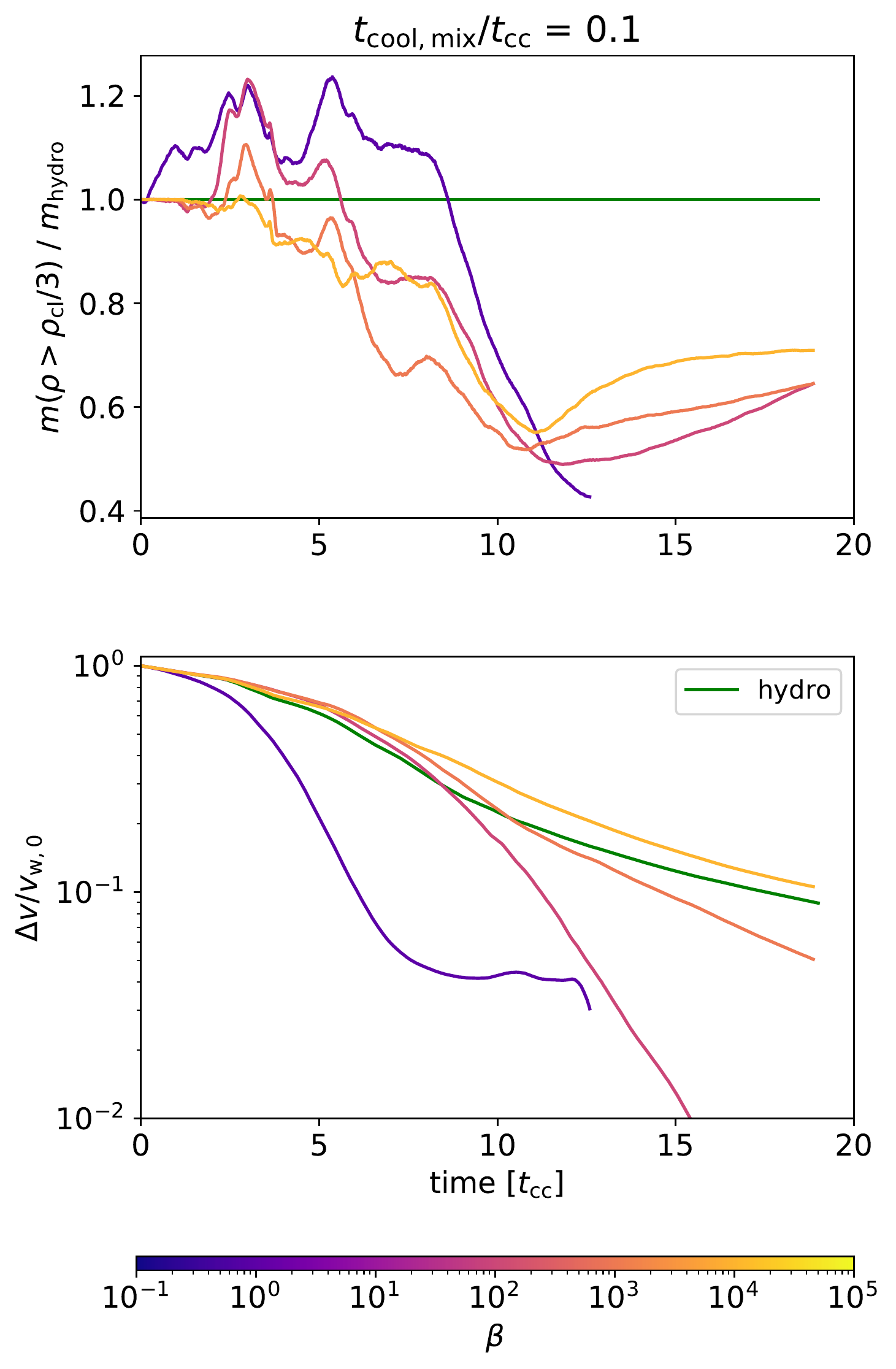}
    \caption{Time evolution of (top panel) the cloud mass normalized to the cloud mass in the hydrodynamic case (hence, the green curve is flat and has a value of unity throughout) and (bottom panel) the relative velocity between the cloud material and hot wind ($\Delta v)$ normalized to the initial wind velocity ($v_{\rm w,0}$).}
    \label{fig:disc:mass_growth_modified}
\end{figure}

The situation is, however, different with more efficient cooling.
In Fig.~\ref{fig:disc:mass_growth_modified}, we show the cloud mass evolution normalized by the hydrodynamic run's mass and additionally show the relative velocity evolution in Fig.~\ref{fig:disc:mass_growth_modified}, for the case of \tratio\ = 0.1.
Even while the $\beta = 10^4$ simulation (orange curve in Fig.~\ref{fig:disc:mass_growth_modified}) is entrained only slightly slower than the hydrodynamic case, it has a reduced mass compared to the hydrodynamic case, suggesting magnetic fields do suppress mixing, if marginally in these cases.

Figure~\ref{fig:disc:mass_growth_modified}, also shows that for $\beta=1$ the relative velocity not only drops to $\sim$half its initial value but is a factor of $\sim$5-10 lower at 8 cloud crushing times than the hydrodynamic case, which is when clouds were destroyed in the limit of weak cooling cases. Furthermore, at later times ($t\gtrsim 10 t_{\rm cc}$), the entrainment is faster also for $\beta=10$ although this is not expected from draping which should play a negligible role for $\beta \gtrsim 10$ (Eq.~\ref{eq:tdrag_rat}).

\begin{figure}
    \centering
    \includegraphics[width=0.9\columnwidth]{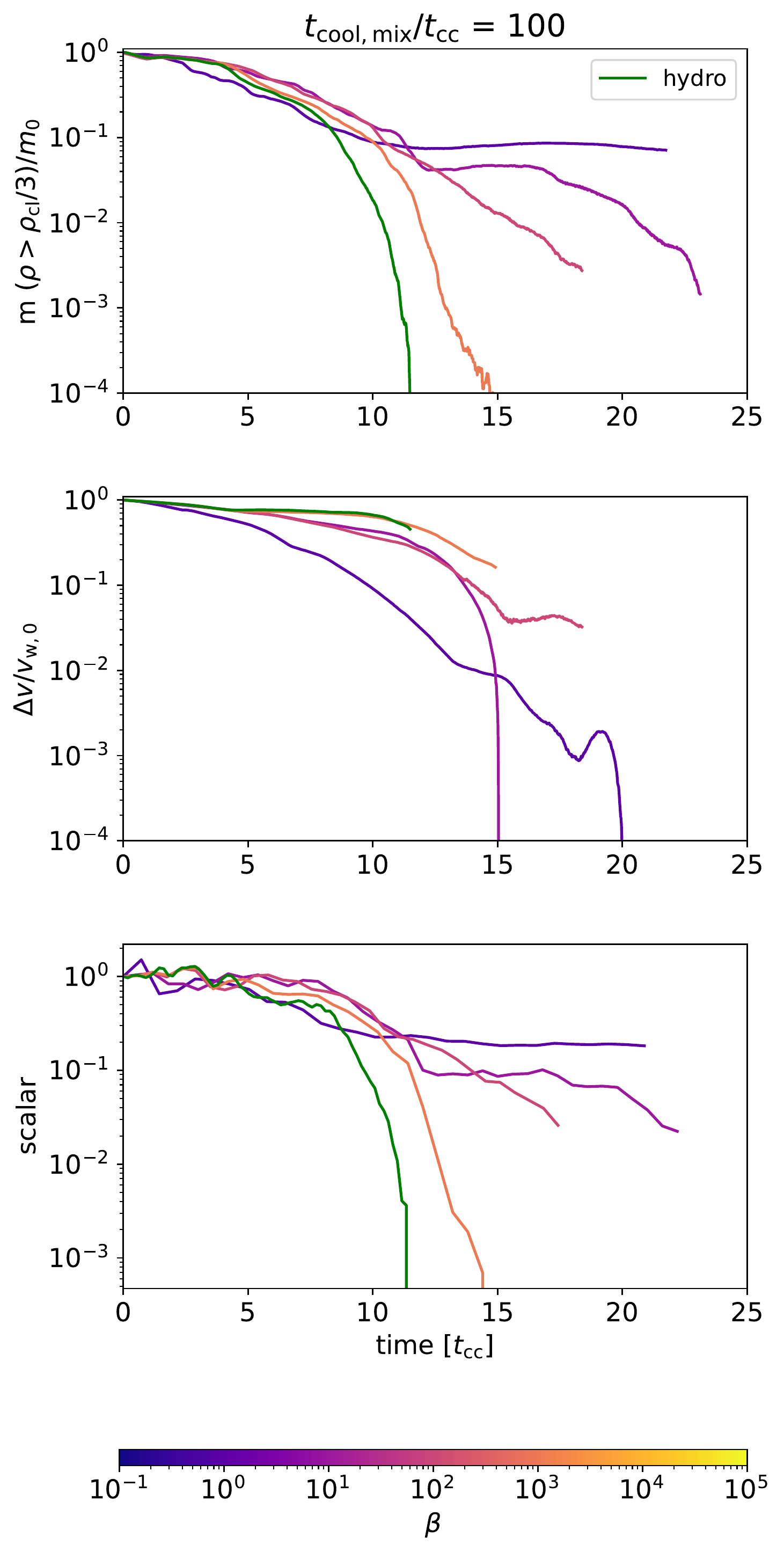}
    \caption{Time evolution of the mass (top panel), relative velocity between the wind and cloud (middle panel), and Lagrangian scalar concentration initially assigned to cold gas (bottom panel) for $t_{\rm cool,mix}/t_{\rm cc} = 100$ as a function of $\beta$ (with the green curve indicating a hydro run for comparison). From this plot, it is clear that lower $\beta$ clouds survive (longer) because they entrain more quickly and mix less / more slowly.}
    \label{fig:disc:scalar}
\end{figure}

To further investigate the ability of magnetic fields to aid cloud survival we show in Fig.~\ref{fig:disc:scalar} time series of cloud mass (top panel), relative velocity (middle panel), and scalar concentration (bottom panel, see Methods~\ref{subsec:scalar}) for \tratio\ = 100, since with this cooling efficiency only the strongly magnetized case $\beta = 1$ survives. 
Even with magnetic fields as weak as $\beta = 10^3$, the lifetime of the cloud is extended by a few cloud crushing times and lower $\beta$ cases take about twice as long as the hydrodynamic case before they are destroyed.
Although $\beta = 1$ and $10$ are completely entrained, with $\Delta v/v_{\rm w,0} < 10^{-4}$, they still have substantial values of their scalar fields, demonstrating entrainment occurs in these cases before the original cloud material is fully mixed.

To summarize: Why does the cold gas survival criterion change by orders of magnitude although the individual MHD effects are rather small? First, recall that we only require $t_{\rm drag}\lesssim \chi^{1/2} t_{\rm cc}$, that is, nominally a reduction of $t_{\rm drag}$ by a small amount for $\chi \sim 100$ to allow for $r_{\rm crit}\rightarrow 0$ (recall $t_{\rm drag}$ scales linearly with $\chi$ so changing $\chi$ by e.g., 4 reduces the drag time by a factor of 4, but the cloud crushing time by only a factor of 2). As mentioned above, \citet{Gronke2020Cloudy} found a critical $\chi \sim 30$ below which clouds were entrained before they were destroyed in even adiabatic simulations.
This implies a disproportionally large change in the critical $t_{\rm cool,mix}/t_{\rm cc}$-criterion is possible for a small change in $t_{\rm drag}$ for $\chi = 100$.

\begin{figure}
        \includegraphics[width=\columnwidth]{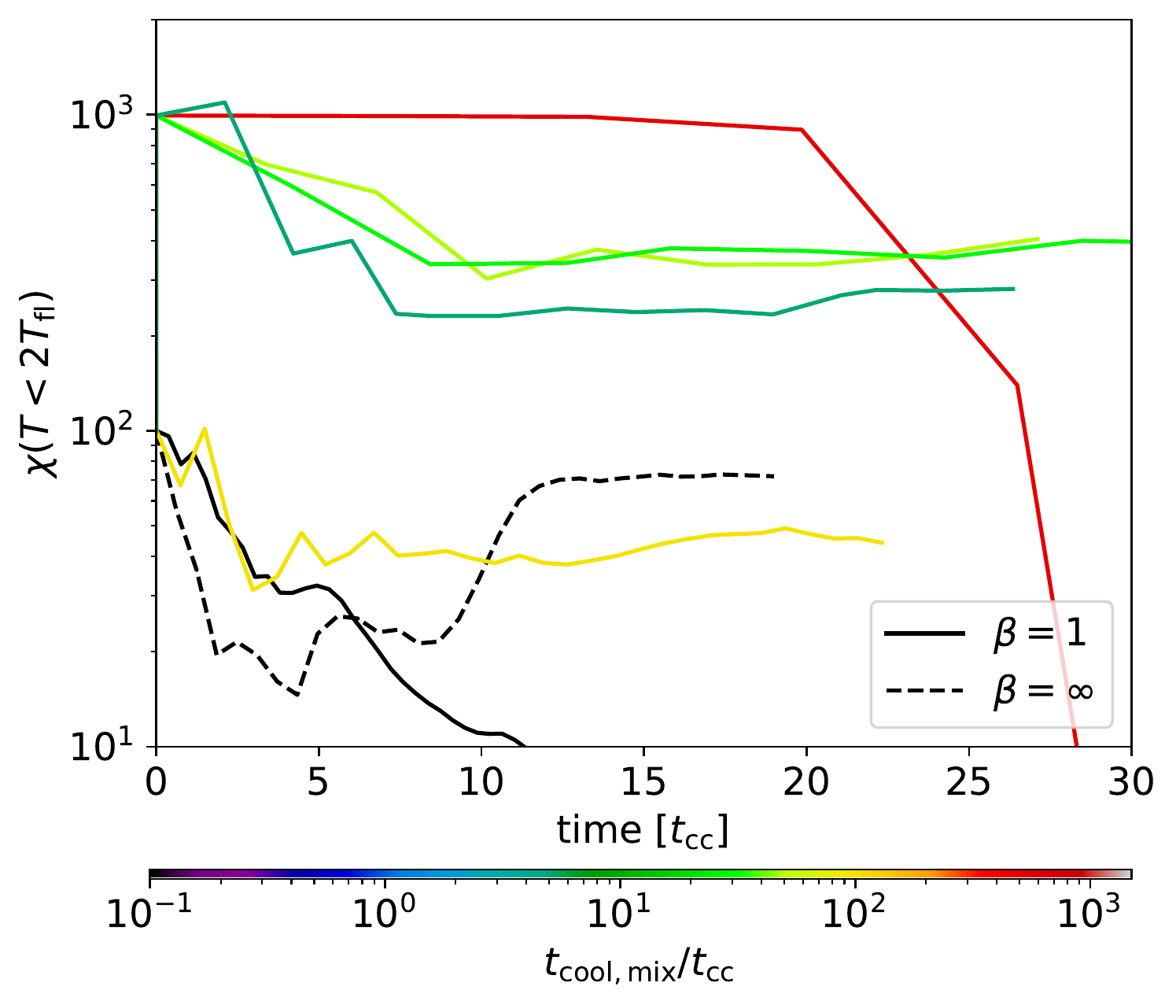}
    \caption{Time evolution of the density ratio 
    $\chi$ between hot tenuous wind and cold dense `cloud' material. In this case we use $T < 2 T_{\rm fl}$ to define `cloud' 
    material but note that the result is very similar for $\rho > \rho_{\rm cl}/3$. The black dashed curve is a hydrodynamic case 
    with $t_{\rm cool,mix}/t_{\rm cc} = 1$ and an initial $\chi = 100$ for reference, and note we include runs with initial $\chi = 100$
    and $\chi = 1000$ as should be evident by their value at time = 0 $t_{\rm cc}$. While the hydrodynamic case loses initial 
    thermal pressure equilibrium, it re-establishes its initial overdensity at $\sim10 t_{\rm cc}$. In contrast, all 
    cases with magnetic fields (except the red curve corresponding to $t_{\rm cool,mix}/t_{\rm cc} = 500$ which is destroyed) maintain a factor of few suppression to their initial overdensity, suggesting magnetic pressure support plays a role in their evolution.}
    \label{fig:disc:overdensity}
\end{figure}

In other words, the question boils down to: Why do magnetic fields cause faster entrainment? First, there is the `draping effect' discussed in \S~\ref{sec:intro}. However, we also demonstrated that with cooling, the entrainment process is faster than expected from draping alone (cf. Fig.~\ref{fig:scatter}). 
To understand this, recall that the drag time $t_{\rm drag}\propto \chi$. Therefore, if magnetic pressure supports more tenuous material than would be stable in thermal pressure equilibrium, then acceleration can occur more rapidly.
In Fig.~\ref{fig:disc:overdensity} we show the time evolution of the overdensity $\chi$, including simulations with $\chi = 10^3$. The case that clearly survives (\tratio\ = 5, the darkest green curve) has the largest decrement in $\chi$ by a factor of $\sim$5, whereas somewhat more inefficient cooling cases (\tratio\ = 30 and 50) only have $\chi$ drop a factor of $\sim$3 and are marginally destroyed or marginally survive (in the \tratio\ = 50 case, the cloud mass drops slightly below 1\% just as the relative velocity is dropping below 1\% after which it might grow at the end of the simulation). The $\chi = 10^3$ case which was clearly destroyed (\tratio\ = 500, red curve) interestingly does not drop in overdensity until it is completely mixed away. This suggests more efficient cooling entrains magnetic fields more effectively which are amplified as they are compressed, providing nonthermal pressure support and reducing the average overdensity. Note that while the hydrodynamic case regains thermal pressure equilibrium and $\chi$ returns to its initial value, the average overdensity remains lower in the cases with magnetic fields.
Hence, the entrainment process even for larger $\beta$ is accelerated -- more than expected from pure `draping'. This leads to comoving (and thus surviving) cold gas and is therefore the main effect for the change in survival criterion found. This point is further discussed in Appendix \S~\ref{app:internalbeta}.

\subsection{Comparison to Previous Work}
We next discuss how our results compare to previous work. One of our most staggering results is the finding that significantly smaller clouds $t_{\rm cool,mix}/t_{\rm cc} \sim 100$ can survive under conditions of low $\beta$ compared to $t_{\rm cool,mix}/t_{\rm cc} \sim 1$ in the hydrodynamical case. 

While many previous studies \citep[e.g.,][]{gregori1999enhanced,banda2016filament,Gronnow2018} included magnetic fields in wind-tunnel simulations, there has been no consensus of whether magnetic fields aid survival and under which conditions. 
\citet{gregori1999enhanced}, for instance, concluded that magnetic fields lead to a faster \textit{destruction} due to faster acceleration and, thus, a shorter Rayleigh-Taylor timescale. 
In contrast, \citet{McCourt2015} show that cold gas can survive when including magnetic fields. 
However, they focus on moderate overdensitites of $\chi\sim 50$ and include radiative cooling. 
Since \citet{McCourt2015} include radiative cooling, it is possible that their simulations effectively fall under the critical $\chi \lesssim 30$ (cf. Eq.~\ref{eq:chi_crit}).

\citet{Li2020} carried out a large suite of simulations which also include magnetic fields. They found an alternative survival criterion to \citet{Gronke2018} used here which compared the hot gas cooling time $t_{\rm cool, hot}$ to an empirically calibrated survival time $t_{\rm life}$ \citep[see][for a discussion and comparison of the criteria in the hydrodynamic case]{Kanjilal2021}. 
Interestingly, the magnetic field does not enter $t_{\rm life}$ and thus the survival criterion of \citet{Li2020} appears at odds with our findings. However, they focus mainly on the $\beta = 10^6$ parameter space where we find the effect to be negligible. Noteworthy, \citet{Li2020} additionally included anisotropic conduction and viscosity which complicates comparison.

\citep{Sparre2020} perform cloud crushing simulations with magnetic fields focusing on the $\beta \geq 10$ case, and in particular on the hydrodynamic case. They do find enhanced cloud survival including magnetic fields -- for example, their $R_{\rm cl} = 15$\,pc cloud survives in their Figure A2 including magnetic fields whereas the same cloud is destroyed in their hydrodynamic run, shown in their Figure D1 -- without discussing the effects of $B$-fields systematically.

\citet{cottle2020launching} performed three-dimensional cloud-crushing simulations with $\chi = 10^3$, $\mathcal{M} = 3.5$, and $\beta = 10$ (and two cases with $\beta = 1$), comparing wind axis-aligned magnetic fields to transverse magnetic fields. 
They found shock-aligned magnetic fields increase the mixing rate by a factor of a few whereas transverse fields drape and pinch clouds leading to a wider perpendicular distribution of cloud material and rapid mass loss. 
Interestingly, \citet{cottle2020launching} claim destruction for their simulations even in a very efficiently cooling regime $t_{\rm cool}/t_{\rm cc} = 10^{-6}$, although presumably they evaluated the cooling time at $T_{\rm cl}$ since we find for their parameters $t_{\rm cool,mix}/t_{\rm cc} \sim 0.1$. 
Also note they use the \citet{wiersma2009effect} cooling curve whereas we use \citet{Sutherland1993} and thus their cooling rate is $\sim$2 lower at $T_{\rm mix}$ and perhaps more significantly their integrated cooling rate below $T_{\rm mix}$ may be substantially reduced compared to ours (see \citealt{wiersma2009effect}, their Figure 1 for a comparison of the cooling curves; also interestingly, their integrated cooling time is longer than ours above $T_{\rm mix}$). 
Nevertheless, one might expect a different outcome in their case since their $\mathcal{M} = 3.5$ is larger than the transonic case studied here.

We do not attempt a full comparison but conclude by noting we consider our studies complementary since we study the transonic case $\mathcal{M} = 1.5$ whereas \citet{cottle2020launching} consider the supersonic scenario $\mathcal{M} = 3.5$.

Regarding morphology of the cold gas, our findings agree with previous studies. For instance, in agreement with \citet{ruszkowski2014impact} we find hydrodynamic simulations have more discontinuous cores whereas magnetized wind-tunnel simulations have a more filamentary, contiguous morphology \citep[among many others, also see][]{tonnesen2010tail,tonnesen2014ties,Jung2022}.

\subsection{Implications for Observations}
Both absorption and emission measurements commonly detect cold $10^4$\,K gas in the circumgalactic medium of objects with virial temperature being $T_{\rm vir} \sim GM/R \sim 10^6$\,K as confirmed in soft X-ray observations \citep{tumlinson2017circumgalactic} (although there may additionally be a $10^7$\,K phase; \citealt{gupta2021supervirial}).
A wide variety of circumgalactic medium observations suggest this cold gas has a scale as small as $\sim$pc (as reviewed in \citealt{McCourt2018}). In addition, detections of rapidly outflowing cold gas in galactic winds close to the expected launching radius of the wind suggest rapid acceleration \citep{Veilleux2020}, and we find magnetic fields induce rapid acceleration even for clouds that are destroyed (recall Fig.~\ref{fig:disc:weakCooling}).

Our results of magnetic-draping enhanced acceleration and low $\beta$ enhanced survival of cold gas shifts the previous survival criterion -- and thus the size of survivable clouds -- by orders of magnitude. This may thus help to explain observations of rapidly outflowing cold clouds and small-scale cold gas in the circumgalactic medium. Moreover, our findings suggest that if the accelerated clouds have dynamically important magnetic fields, observations should seek sub-pc clouds, which may be discovered by future deep H{\small{I}} observations such as with ASKAP \citep{dickey2013gaskap}, JVLA \citep{murthy2021h}, MeerKAT \citep{pourtsidou2017hi}, or other observatories.

Moreover, our simulations may help to explain the cold gas observed in the tails of jellyfish galaxies that form when late-type spirals are subjected to ram pressure from high-pressure intracluster media (most spectacularly, the $\sim$60\,kpc star-forming tail of D100 in Coma, \citealt{cramer2019spectacular} but see \citealt{poggianti2017gasp} for additional examples from the GASP survey). Simulations of jellyfish galaxies find that dense gas is difficult to strip \citep{tonnesen2012star} but stripped gas clouds may evolve due to mixing with the ambient intracluster medium \citep{tonnesen2010tail,tonnesen2021s}.  Our results suggest including magnetic fields in future simulations of jellyfish galaxies may allow smaller clouds to grow.

Another important observational implication is the strong magnetic pressure support and the consequent lower overdensitites found (cf. Fig.~\ref{fig:disc:overdensity}). This effect might explain cold gas found to be out of (thermal) pressure equilibrium in the CGM \citep{werk2014cos}.

\subsection{Caveats}
\begin{itemize}
    \item \textbf{Resolution}: fiducial values are set to a fixed resolution $r_{\rm cl}/d_{\rm cell} = 16$
    to permit a wide parameter sweep of $\beta$ and $t_{\rm cool,mix}/t_{\rm cc}$. Previous simulations find this resolution is adequate for converged mass growth \citep{Gronke2020Cloudy,tan2021radiative}. A test of numerical convergence is shown in Appendix \S~\ref{sec:numericalconv}.

    \item \textbf{Numerical limitations}: 
    The cloud-tracking system aims to avoid gas outflows of the simulation domain, enabling the use of a reduced 3D box-size. 
    Nevertheless, $\beta = 1$ simulations exhibit a substantial expansion of cloud material orthogonal to the wind axis (as seen by several previous studies such as \citealt{cottle2020launching}) which required larger box sizes and hence costlier simulations. Additionally, $\beta = 1$ and $\chi = 10^3$ simulations were numerically unstable at late times (omitted from our analysis) which made it difficult to definitively assess whether borderline clouds would end up being destroyed or survive if we could stably evolve our simulations longer. Future simulations utilizing a more stable Riemann solver such as HLL3R \citep{waagan2011robust} or otherwise more stable numerical methods may help us study lower $\beta$ and higher $\chi$ evolved to longer times.
    \item \textbf{Initial parameters}: we performed our simulations primarily at $\chi = 100$ yet clouds likely span overdensities of $10^{2-4}$ (that is, they exist in pressure equilibrium, such as $10^4$\,K gas in the $10^8$\,K intracluster medium). Note however that higher overdensities are more numerically expensive and we wished to run a wide sampling of simulations to elucidate clearly the impact of radiative cooling and magnetic fields on cloud survival. Moreover, we only ran simulations at $\beta_{\rm wind} = \beta_{\rm cl}$ whereas in reality these values may likely be different. However, \citet{Gronke2020Cloudy} studied independently varying $\beta_{\rm cl}$ and $\beta_{\rm wind}$ and found $\beta_{\rm wind}$ was the most determining parameter, so we do not expect significant modifications to our results.
    Furthermore, we only explored transonic $\mathcal{M}=1.5$ winds. Such are expected to be relevant to the conditions of the launching radius of galactic winds \citep{Chevalier1985}; however, if clouds are sourced from the CGM at greater distances from the base of the galactic wind, exploring higher Mach numbers would be more relevant. We leave such exploration to future work and refer the reader to \citet{Sparre2020} for the case with $\beta = 10$ who studied $\mathcal{M} = 4.5, 1.5$ and 0.5 and \citet{cottle2020launching} for cases of $\beta = 10$ and 1 with $\mathcal{M} = 3.5$.

   \item \textbf{Interplay of complex dynamical processes}: 
   Pressure-driven fragmentation of clouds undergoing thermal instability as they are subjected to a hot wind has not been studied in this work, but would be an interesting avenue for future research. Furthermore, we neglected potentially important physical processes such as conduction, viscosity, turbulence, and cosmic rays which may impact our results. We plan to study these effects in future work. Recently,  \citet{Bruggenprep} have conducted a study focusing on the problem of thermal conduction and its impact on cloud survival in combination with magnetic fields.

\end{itemize}

\section{Conclusions.}
\label{sec:conclusions}
We perform simulations of radiative cold magnetized clouds subject to hot magnetized winds. We ran simulations with $\beta = 1, 10, 100, 10^4,$ and $10^{20}$ (hydrodynamic limit) each with various \tratio\ from 0.1 to $10^4$. 
Introducing strong magnetic fields in such plasmas favours their survival for a critical \tratio\ 100 fold above the hydrodynamical scenario in conditions of low $\beta \sim 1$ and 10 fold for $\beta \sim 10$ (cf. Eqns.~\ref{eq:tratio_fit} \& \ref{eq:rcrit_fit} for a cold gas survival criterion in a magnetized wind). For $\beta \gtrsim 10^4$ we recover the hydrodynamical survival criterion.

As most astrophysical plasmas are magnetized, this implies that much smaller clouds can survive ram pressure acceleration than previously thought.
We attribute this strong impact to a more rapid entrainment process leading to an entrainment time shorter than the cloud's destruction time. 
We find that this rapid entrainment cannot be explained by magnetic draping alone but is also due to a lower cold gas overdensity (and hence faster acceleration since $t_{\rm drag}\propto\chi$) caused by non-thermal pressure support because of compressed magnetic fields.
In summary, we find that the key aspect is the \textit{interplay} of magnetic field lines and cooling which has a combined much stronger effect than the individual components themselves.

Other results of our work are in broad agreement with previous publications on cold magnetized cloud survival that generally find enhanced cloud survival when including radiative cooling and magnetic fields, as well as simulations that find enhanced destruction in adiabatic simulations with magnetic fields.
We find magnetic fields do not completely arrest Kelvin-Helmholtz instability except possibly at late times in cases of efficient cooling and $\beta = 1$ as well as cases with weak but non-negligible cooling $t_{\rm cool,mix} / t_{\rm cc} = 100$ with $\beta = 1$.

The rates of cold mass growth in MHD scenarios appear to coincide with past and simulated hydrodynamical cases for the same initial parameters, reaching constant stable growth at late stages, except a slight suppression of growth for $\beta = 1$.

While our simulations help to understand the evolution and survival of cold gas in hot winds, 
our results have certain limitations related to the restricted range of parameters we explored.
Supersonic Mach numbers, higher overdensities, turbulence, conduction, viscosity, cosmic rays, and perhaps radiation pressure may have an impact on our results. In the future, we would like to extend this work for a larger parameter range to determine the impact this may have on the now more-advanced, classical cloud survival problem. 



\section*{Acknowledgements}
We thank the referee for their useful comments and feedback.
The authors thank the MPA internship program, which made this research possible. We also thank Peng Oh, Evan Scannapieco, Marcus Brüggen and Stephanie Tonnesen for helpful conversations and Hitesh Kishore Das for help with visualisations.
MG thanks the Max Planck Society for support through the Max Planck Research Group.
The simulations and analysis in this work were and are supported by the
Max Planck Computing and Data Facility (MPCDF) computer cluster Freya. This project utilized the visualization
and data analysis package yt \citet{Turk2011}; we are grateful to
the yt community for their support.
This research has made use of NASA's Astrophysics Data System, matplotlib, a Python library for publication quality graphics \citep{Hunter:2007}, and NumPy \citep{harris2020array}.
The authors gratefully acknowledge the Gauss Centre for Supercomputing e.V. (www.gauss-centre.eu) for funding this project by providing computing time on the GCS Supercomputer SuperMUC-NG at Leibniz Supercomputing Centre (www.lrz.de).

\section*{Data Availability}
Data to confirm the results of this work will be shared upon reasonable request to the authors.
 



\bibliographystyle{mnras}
\bibliography{paper} 




\appendix

\section{Numerical convergence}
\label{sec:numericalconv}
Previous work has shown that even at resolution as low as 8 cell lengths per cloud radius leads to convergence in mass evolution and growth \cite{Gronke2020Cloudy}. In short, this is because the mixing time $t_{\rm eddy}\propto l^{2/3}$, i.e., increases with scale. In other words, the factor limiting mixing is the large eddies which we are resolving \citep{tan2021radiative}. For our systematic study of magnetised plasmas, we employ a cell length/cloud radius ratio of 16. Using the fiducial run of initial values $\beta = 1$, $t_{\rm cool,mix}/t_{\rm cc}$ = 0.1, we contrast our simulation with the produced mass evolution for a cold cloud of resolutions $r_{\rm cl}/d_{\rm cell} = 32$.

\begin{figure}
    \centering
    \includegraphics[width=\columnwidth]{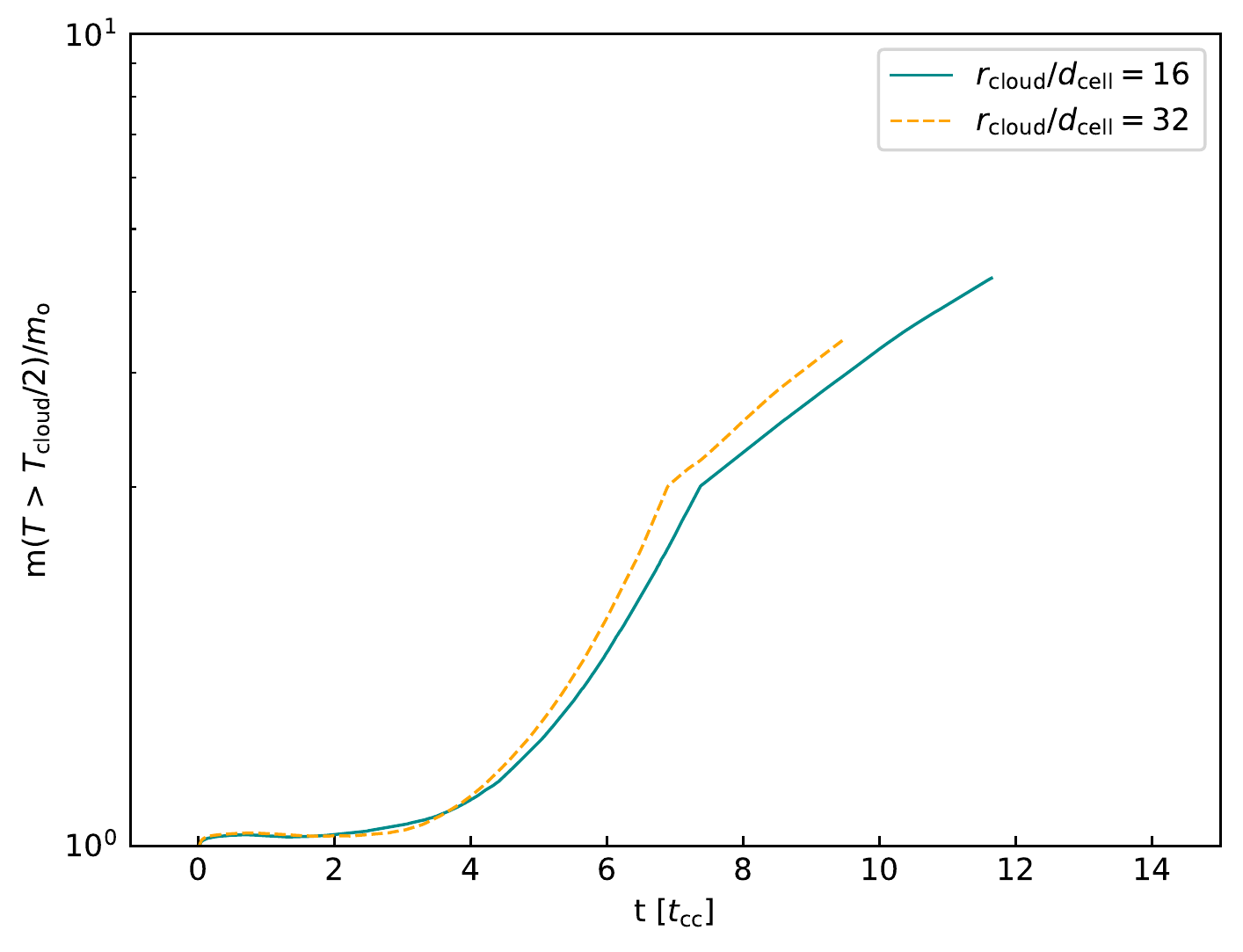}
    \caption{Runs for initial conditions $\beta = 1$, $t_{\rm cool,mix}/t_{\rm cc}$ = 0.1 employing resolutions of $r_{\rm cl}/d_{\rm cell} = 32$ (dashed-orange) and $r_{\rm cl}/d_{\rm cell} = 16$ (solid-blue), respectively. Further comparative analysis for different resolutions can be compared in \citet{Gronke2020Cloudy}.}
    \label{fig:conv_test}
\end{figure}

These two resolutions displayed in \ref{fig:conv_test}, represented by the solid-blue and dashed-orange lines, respectively, exhibit only minute deviations of order $\mathcal{O}{(}{-1)}$ at $\sim$7\,$t_{\rm cc}$, demonstrating similar behaviour for both resolutions.

\section{Shear evolution}
\label{app:entrainmentTime}
\begin{figure}
    \centering
    \includegraphics[width=\columnwidth]{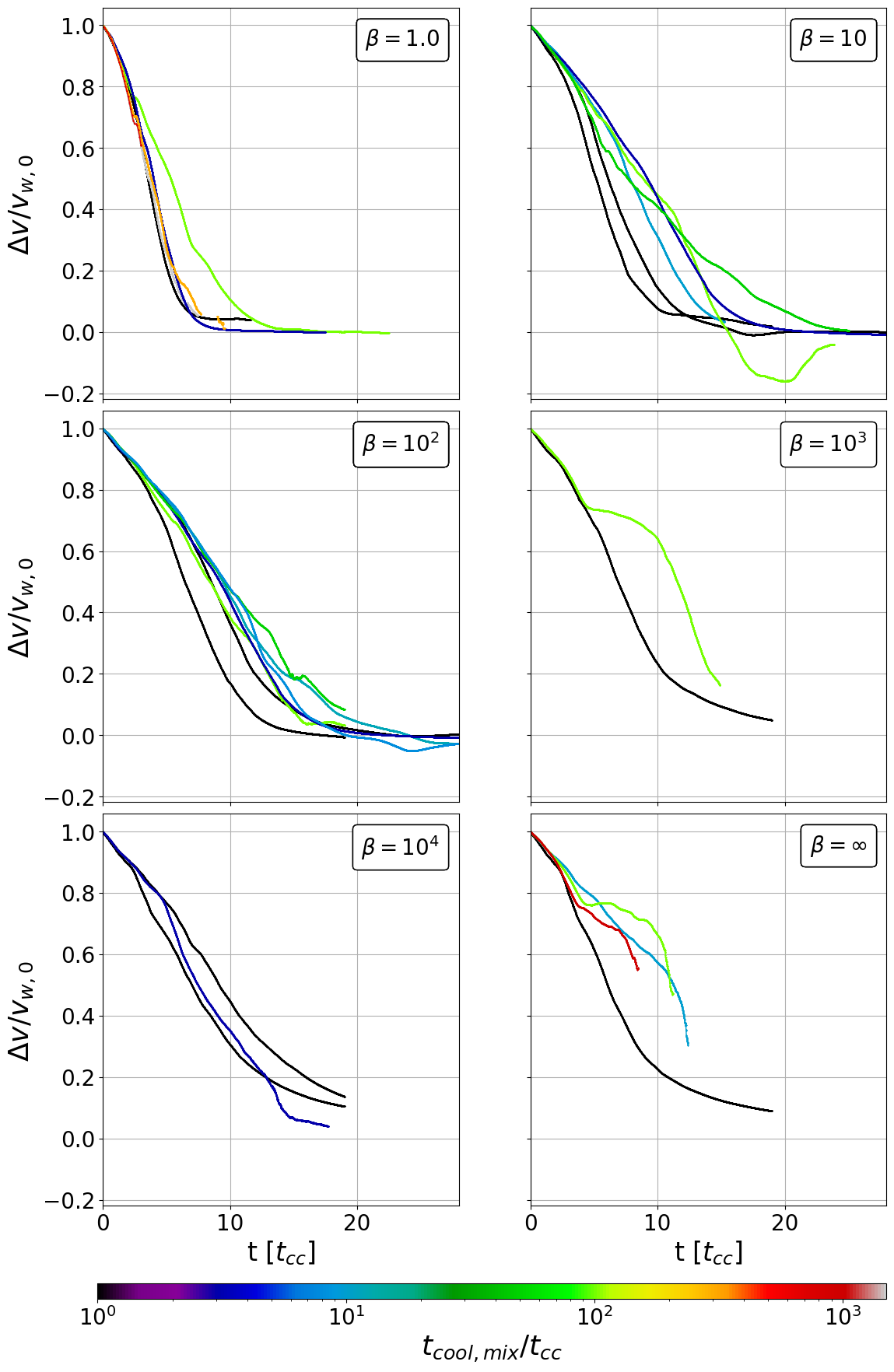}
    \caption{ Evolution of the shear between cold gas and a $ \mathcal{M} \sim 1.5$ hot wind with an overdensity of $\chi \sim 100$. We vary $\beta$ (from  $\beta \sim 1$ in the upper left
to  $\beta = \infty$ in the lower right panel) and cooling strength (indicated by
the line color).}
    \label{fig:vt_detailed}
\end{figure}

 The entrainment of cold gas through momentum transfer from the hot wind to the cloud is an important element in solving the cloud crushing problem. 
 Figure~\ref{fig:vt_detailed} shows the evolution of the velocity difference between the hot and the cold medium for a variety of our runs.

In the y-axis, we plot the shear velocity of the cloud with respect to the wind, normalised to the initial wind speed and thus 1.0 represents a static cloud and 0 indicates full entrainment by the wind. 
The evolution of velocities is displayed in cloud crushing timescales, $t_{\rm cc}$. Cooling efficiencies are encoded by the bottom colourbar. Significant variations in curve steepness can only be seen for low betas, more specifically, for $\beta = 1$.

\begin{figure}
    \centering
    \includegraphics[width=0.8\columnwidth]{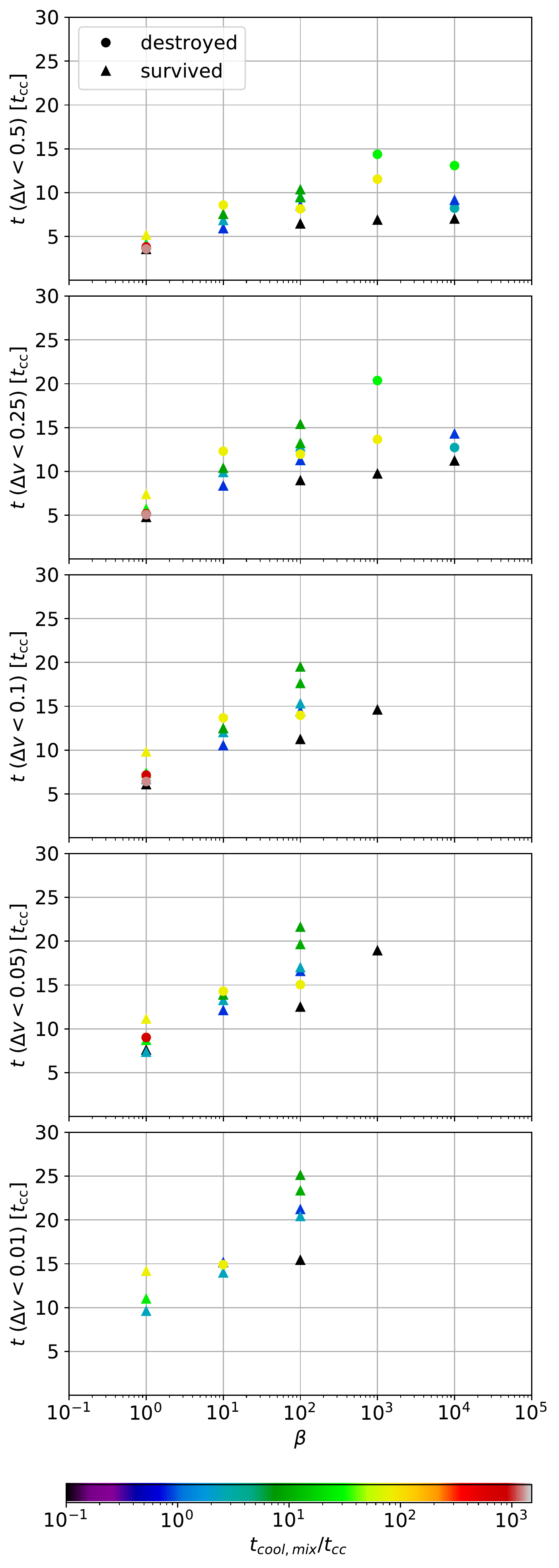}
    \caption{Time for cloud material to be `entrained' ($\Delta v < \alpha$ for thresholds from $\alpha = 0.5$ (top) to $\alpha = 0.01$ (bottom)). Note that in many of our simulations, clouds are completely destroyed or are already evidently growing (at which point we stop restarting them) by $\Delta v < 0.01$ or even $\Delta v < 0.1$. One would expect that clouds that survive entrain faster than clouds that are destroyed. However, we observe several exceptions to this expected trend, particularly at $\beta = 1$ in which case $t_{\rm cool,mix}/t_{\rm cc} = 300$ (red circle, which is destroyed) entrains even to $\Delta v < 0.05$ faster than $t_{\rm cool,mix}/t_{\rm cc} = 100$ (yellow triangle, which survives). However, $t_{\rm cool,mix}/t_{\rm cc} = 300$ is destroyed before entraining to $\Delta v < 0.01$ suggesting the last 10\% of the entrainment (which requires 1.5 - 2.5 $t_{\rm drag}$ is critical for cloud survival.}
    \label{fig:app:entrainmentTime}
\end{figure}

Figure~\ref{fig:app:entrainmentTime} shows multiple versions of figure~\ref{fig:scatter} with varying "entrainment time", defined as the time when the shear velocity drops below a certain threshold. A faster rate in cloud acceleration appears to be related to faster cooling efficiencies for the clouds (at fixed $\beta$), expressed in terms of $t_{\rm cool,mix}/t_{\rm cc}$ and which is mapped accordingly to the bottom colourbar. Overall, moving to the right in the x-axis displays a $\sim$ few fold increase in entrainment time as a function of $\beta$, recovering the fiducial timescales from figure~\ref{fig:scatter} as we move upwards to the top subplot.    

Note that, although destroyed clouds might be expected to entrain later than survived clouds, some datapoints reveal the opposite (e.g. $t_{\rm cool,mix}/t_{\rm cc}=300$ for $\beta =1$). However, we observe the cold gas for these cases to get destroyed before entraining to lower shear values (e.g. $\Delta v < 0.01$), which suggests that the final stages of entrainment can be decisive in the final fate of the cloud.

\section{Turbulence suppression}
\label{app:turbulence}
\begin{figure}
    \centering
    \includegraphics[width=\columnwidth]{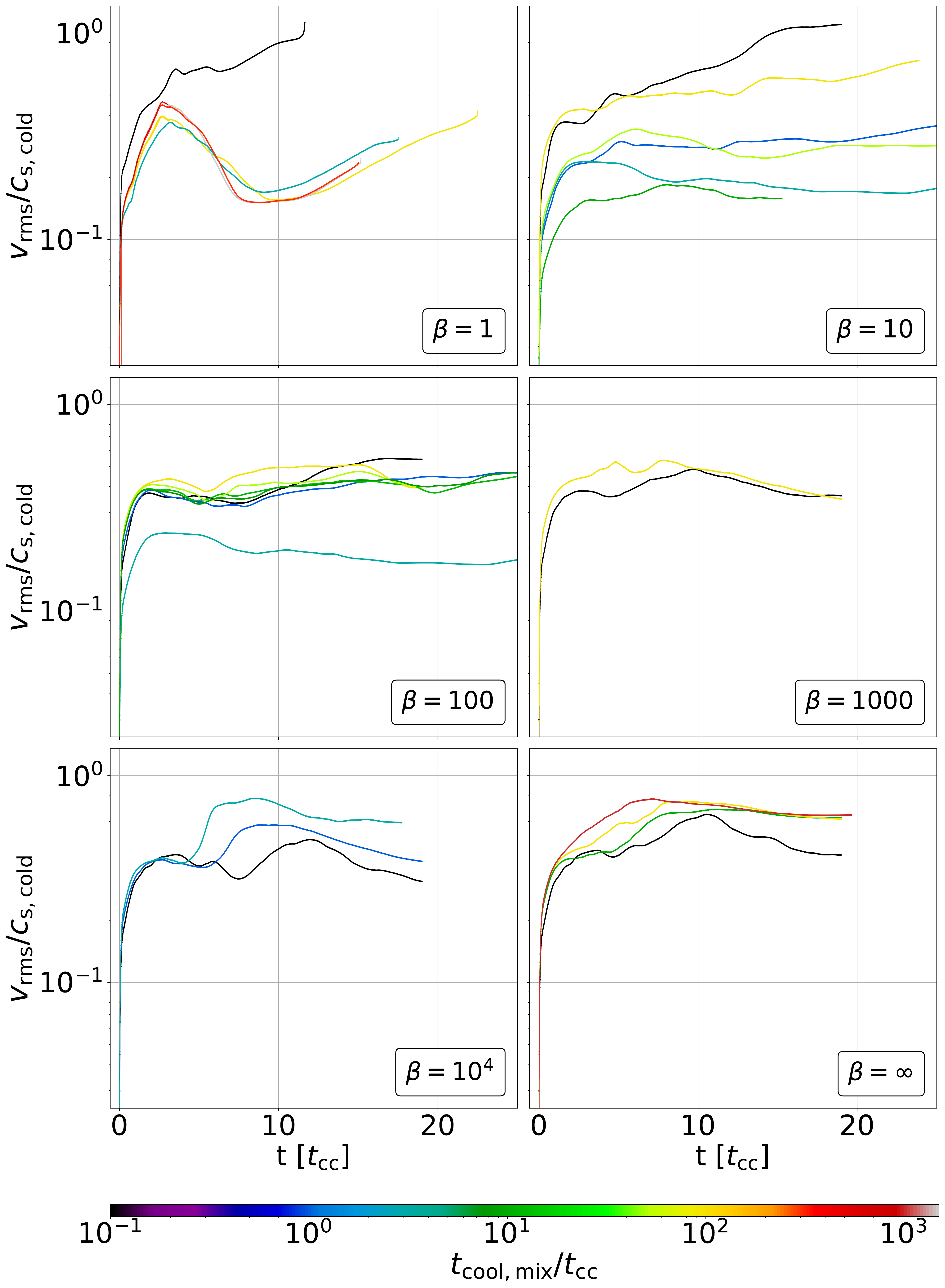}
    \caption{Mass-weighted root-mean-square velocity of the speed components orthogonal to the wind. As discussed in \S~\ref{sec:disc_survival} and Appendix~\ref{app:turbulence}, magnetic fields have a small effect on the turbulent velocity.}
    \label{fig:app:turbulentflow}
\end{figure}

The influence of magnetic fields on hydrodynamical instabilities such as Rayleigh-Taylor or Kelvin-Helmholtz is an on-going topic of discussion in the literature (e.g., \citealp{bubbles,Fielding2020}). In figure ~\ref{fig:app:turbulentflow}, we study this problem in more detail by showing the evolution of the mass-weighted, geometric-mean velocity of cold gas orthogonal to the direction of the wind. The velocity is expressed as a fraction of the sound speed of gas at $T_{\rm cl}$, i.e. $v_{\rm rms}/c_{\rm s, cold}$, where $c_{\rm s, cold}$ represents the sound speed of the cold phase.

Broadly, the speed decreases slightly once the wind encounters the cloud, and only exhibits a small deviation in late-time behaviour with respect to the general trend for the case of $\beta = 1$.
As discussed in \S~\ref{sec:discussion}, simulations experience similar features in turbulence both under the presence of magnetised plasmas and without them.
On average, we observe that late time $v_{\rm rms}/c_{\rm s, cold}$ tend to unity, only varying by a factor of $\sim$ few. Such a level of late time turbulence is expected from pulsations of the cloud; that is, we previously found late time turbulence is not shear-driven \citep{Gronke2020Cloudy,gronke2023,abruzzo2022taming}.

\section{Internal plasma beta evolution}
\label{app:internalbeta}

\begin{figure}
    \centering
    \includegraphics[width=\columnwidth]{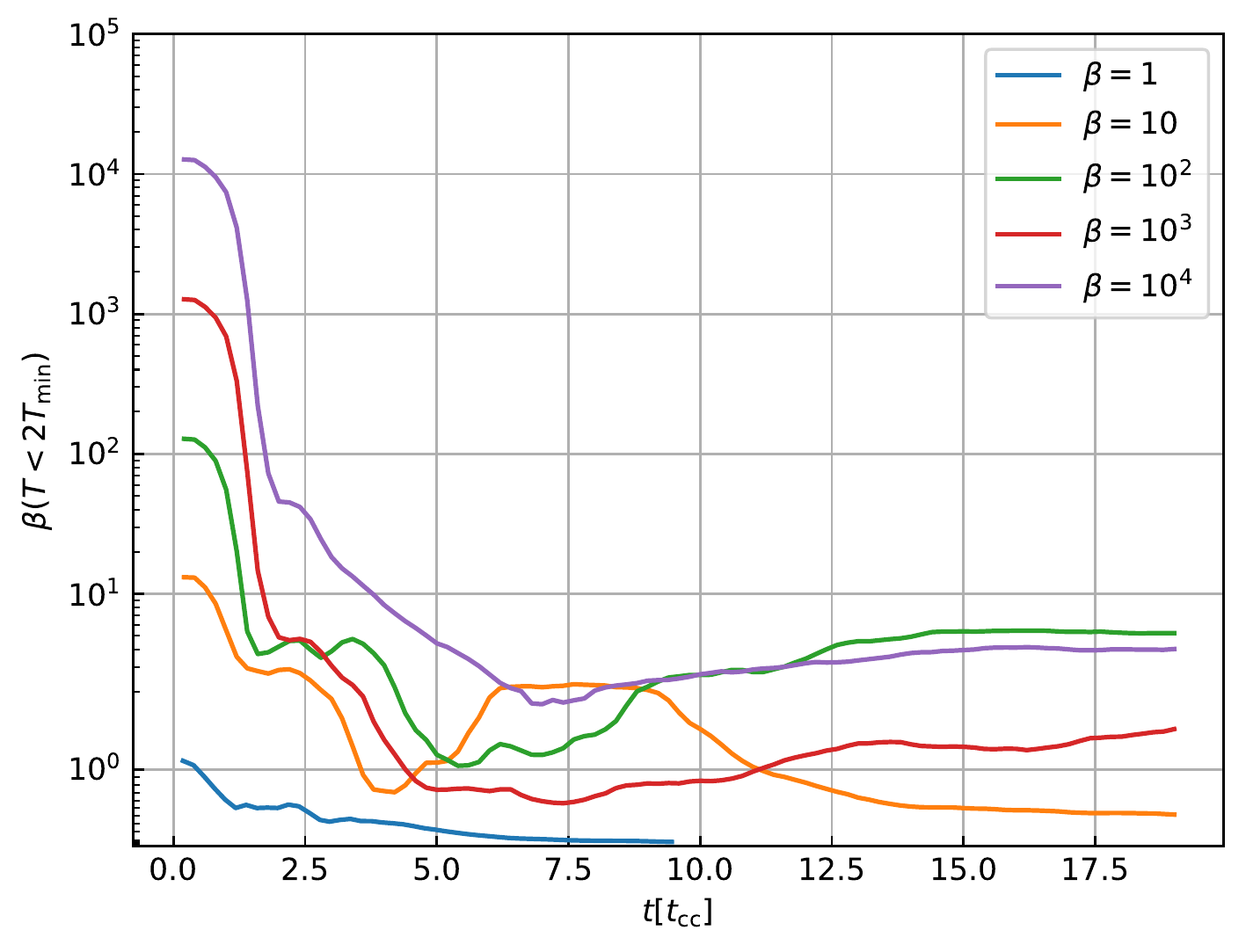}
    \caption{Evolution of plasma beta, $\beta$, for runs of $\tratio = 0.1$ and $\beta = 10^{4}$, $\beta = 10^{3}$, $\beta = 10^{2}$, $\beta = 10$, $\beta = 1$, represented by red, green, orange and blue solid lines, respectively. Curves kick off from their predetermined initial pressure to thermal ratio values and share a subsequent decrease. Final values are restricted well below 10, recovering values of beta $\beta \sim $ few for survived overdense clumps within the simulation box. }
    \label{fig:app:intB}
\end{figure}

In \S~\ref{sec:disc_survival} (cf. Fig. ~\ref{fig:disc:overdensity}), we argue that magnetic pressure support accounts for the drop in late-time cloud overdensities in scenarios where $\beta$ is close to unity. This contrasts with purely hydrodynamical simulations, where overdensities $\chi$ recover their initial values after a few $t_{\rm cc}$. 
Figure \ref{fig:app:intB}, representing the internal magnetic field evolution of cold gas, shows direct evidence of the existence of significant magnetic pressure in the late-time evolution, supporting the hypothesis  that magnetic fields can support this difference in overdensities. Specifically, we computed $\beta = P_{\rm gas}/P_{\rm B}$ via an ideal gas equation of state for $P_{\rm gas}$ and $P_{\rm B} = B^2 / (8\pi)$. We restricted our computation of $\beta$ to cells satisfying $T < 2 T_{\rm min}$ and plot the median value at each time. Notice that $T_{\rm min}$ represents the floor temperature of the simulation.

The plot studies the evolution of plasma beta, $\beta$, for runs of $\tratio = 0.1$ and various initial betas (purple, $\beta = 10^{4}$; red, $\beta = 10^{3}$ ;green, $\beta = 10^{2}$; orange, $\beta = 10$; blue, $\beta = 1$). Our results suggest cold gas exhibits values $\sim$few, regardless of the initial $\beta$.
This is in agreement with previous studies which show a similar enhancement in magnetic field strength in a cooling, multiphase medium \citep[e.g.,][]{Gronke2020Cloudy,das2023magnetic}.

Interestingly, we do find values $\beta < 1$ which indicate compression of the magnetic fields due to cooling as the primary source of field enhancement \citep[see discussion in ][]{das2023magnetic}.

\bsp	
\label{lastpage}
\end{document}